\documentclass[sigconf]{acmart} 

\usepackage{multicol}
\usepackage{soul, xcolor, color}
\usepackage{algorithm}
\usepackage[noend]{algpseudocode}
\usepackage{caption}
\usepackage{subcaption}
\usepackage{balance} 
\usepackage{mathtools}

\newcommand{\new}{\vspace{0.09in}}
\newcommand{\minus}{\vspace{-0.05in}}

\newcommand{\squishlist}{
	\begin{list}{$\bullet$}
		{  \setlength{\leftmargin}{0.8em}
	} }   
	\newcommand{\squishend}{
\end{list}  }

\renewcommand\footnotetextcopyrightpermission[1]{} 
\setcopyright{none}

\acmDOI{}

\acmISBN{}

\acmPrice{}

\usepackage[normalem]{ulem}			
\usepackage[utf8]{inputenc}			
\usepackage{cleveref}
\usepackage{amsmath}
\usepackage{enumitem}
\usepackage{multirow, hhline}
\crefname{section}{§}{§§}
\Crefname{section}{§}{§§}

\graphicspath{{figs/}}

\begin{document}

\newcommand{\name}{{\sc CoDir}}

\renewcommand{\hl}[1]{#1}
\title{\vspace{-0.15in} Inferring Facing Direction from Voice Signals \vspace{-0.05in}}
\author{Yu-lin Wei*}
\affiliation{%
\institution{University of Illinois at Urbana-Champaign}
}

\author{Rui Li*}
\affiliation{%
\institution{Samsung AI Center, Cambridge}
}

\author{Abhinav Mehrotra}
\affiliation{%
\institution{Samsung AI Center, Cambridge}
}

\author{Romit Roy Choudhury}
\affiliation{%
\institution{University of Illinois at Urbana-Champaign}
}

\author{Nic Lane}
\affiliation{%
\institution{Samsung AI Center, Cambridge}
}
\vspace{0.1in}
\thanks{*Co-primary authors}
\maketitle

\section*{Abstract}
Consider a home or office where multiple devices are running voice assistants (e.g., TVs, lights, ovens, refrigerators, etc.).
A human user turns to a particular device and gives a voice command, such as ``Alexa, can you ...''.
This paper focuses on the problem of detecting which device the user was facing, and therefore, enabling only that device to respond to the command.
Our core intuition emerges from the fact that human voice exhibits a directional radiation pattern, and the orientation of this pattern should influence the signal received at each device. 
Unfortunately, indoor multipath, unknown user location, and unknown voice signals pose as critical hurdles.
Through a new algorithm that estimates the line-of-sight (LoS) power from a given signal, and combined with beamforming and triangulation, we design a functional solution called {\name}.
\hl{Results from $500+$ configurations}, across $5$ rooms and $9$ different users, are encouraging.
\hl{While improvements are necessary}, we believe this is an important step forward in a challenging but urgent problem space.

\section{Introduction}
With rapid advances in speech recognition, conversational AI, and on-device compression, voice assistants will soon arrive as a system-on-chip (SoC) \cite{web:audioSoC, web:qualcomm}.
Market surveys project that by 2021, many IoT appliances in homes and offices -- including lights, fans, faucets, washers, dryers, security cameras, TVs, etc. -- will embed these SoCs \cite{web:moen, web:IHSmarkit}.
Users would be able to talk to these devices and such interactions would become frequent, commonplace.
\new

In this emerging future, an important question pertains to the ambiguity of addressing these IoT devices.
A command like “Alexa, turn off after 5 minutes'' does not identify the device the user is addressing. 
Naming each device and prefacing voice commands with that name is a possibility (e.g., ``kids bedroom floorlamp, turn off after 5 minutes'').
However, such naming conventions can be problematic.
For instance, (1) each light, faucet, TV, or music speaker may need to be named individually (e.g., ``guest bathroom faucet''), making it difficult to remember or coordinate across people \cite{web:Smartthings, web:alexaLights}.
(2) Names given by users will suffer from lower reliability, unlike well-trained wake-words like ``Alexa'' or ``OK Google'' \cite{sainath2015convolutional, trmal2017kaldi}.
(3) Finally, humans do not always call each individual by name before speaking to him/her, hence calling each device by name, every time, can be burdensome.
\new

A desirable solution is one that mimics natural human behavior, i.e., if a user turns to a TV and says ``Alexa, turn off in 5 minutes'', that TV should respond while other devices should ignore the command.
We call this the problem of ``facing direction''. 
Stated generally, {\em if a user faces a specific device and gives a voice command, can all the $N$ devices (that overheard the voice signals) collaboratively identify the device the user was facing.}
\new

\begin{figure}[hbt]
\centering
\includegraphics[width=3.3in]{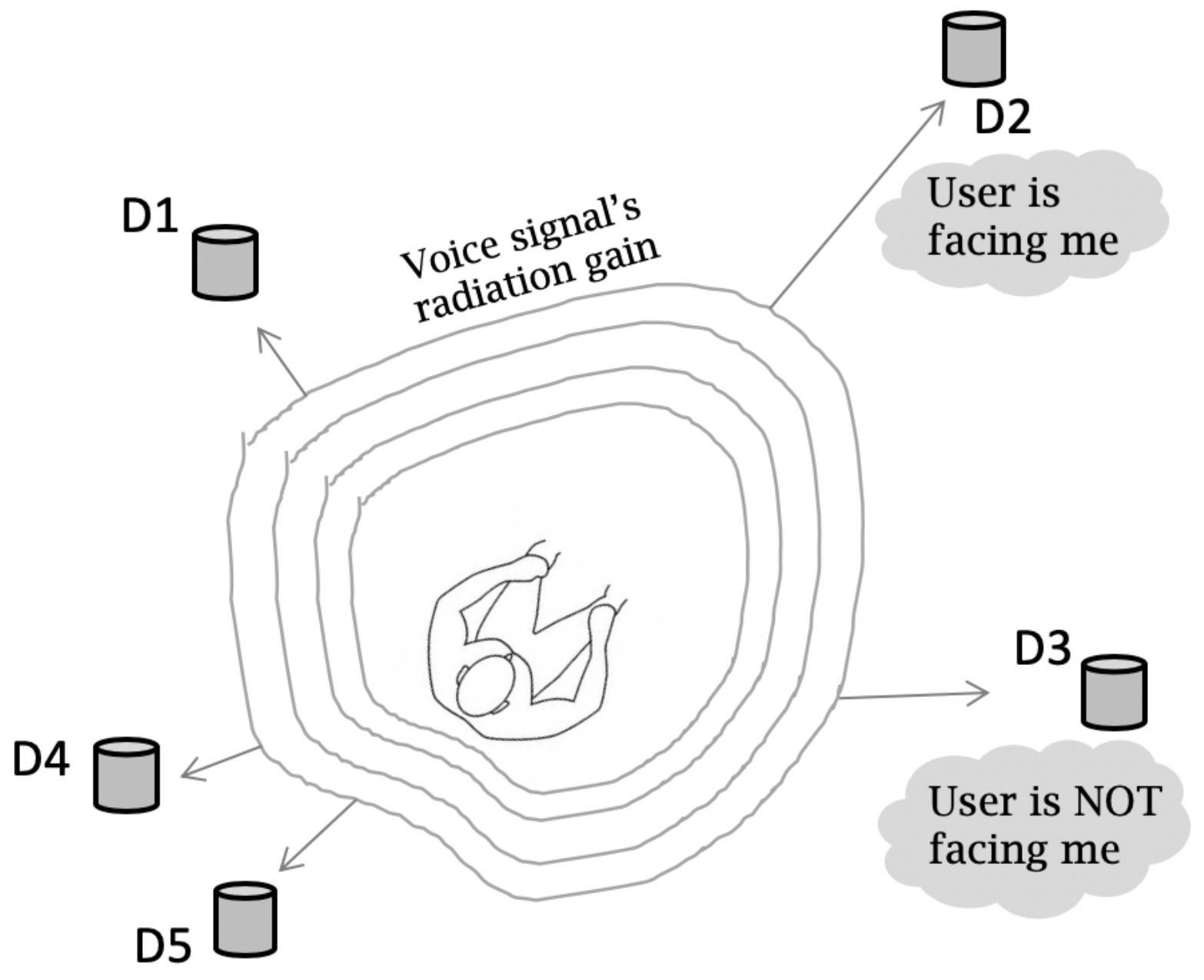}
\minus \minus
\caption{The problem scenario.}
\label{fig:topview} 
\end{figure}

In thinking about solutions to this problem, we make $2$ observations.
(1) If voice radiation from the user's mouth is omnidirectional, then there is no way to infer the facing direction from N overheard signals.
This is because, for a fixed user location, the user can face different devices, and yet the received signals will not change at any of the devices.
Fortunately, human voice radiation is not omnidirectional; instead a larger fraction of energy radiates towards the facing direction compared to the sides and back (toy example in Figure \ref{fig:topview}).
Any solution would need to utilize this information.
\minus  \minus

(2) Collaboration among devices is crucial, i.e., it is hard for any single device to tell whether the user is facing it.
This is because the signal received by a device is not only a function of the user's facing direction, but also the user's location and voice loudness.  
For instance, the signal power recorded at a device can be the same when the user speaks softly facing the device, or speaks loudly facing away from the device. 
Determining the accurate voice power (at the human mouth) is almost infeasible from a single device.
On the other hand, if $N$ devices collaborate and compare their received powers, it may be possible to infer facing direction. 
\new

Given these 2 guidelines (i.e., the need to exploit human radiation pattern and device collaboration), our high level intuition is as follows.
Consider a toy case where the devices are placed in a circle and the user is in the center of this circle.
When the user speaks, the signal-powers recorded across these devices should match the shape of the human's radiation pattern.
Thus, the device with maximum power can be identified as the one the user is facing.
\new

{\em But what happens when the devices are scattered randomly in a real home/office environment?}
Our idea is to eliminate the influence of different distances on each of the devices, so that the received power is only a function of the radiation pattern (like the case of a circle).
For this, we first triangulate the user's location by using the angle of arrival (AoA) at each device\footnote{Let us assume that the device locations are known but we will relax this assumption by also localizing these devices.}.
Once the user location is known, her distance to each device, $d_{i}$, is also known.
The power at each device is then ``equalized'' by multiplying it with a function of $d_{i}$.
Thus, devices that are far away (and record less power) are multiplied with a larger distance, while nearby devices are amplified less.
These equalized powers are now a function of the radiation pattern alone, and the maximum among them should reveal the facing direction.
\new

Not surprisingly, this idea suffers from a fundamental challenge -- multipath.
The equalization idea above is meaningful only when the signal power is derived from the line of sight (LoS).
When multipath pollutes this power, the results of equalization becomes random.
Since the source voice signal is not known, standard channel estimation techniques do not apply (hence the LoS cannot be separated from the multipath).
Hence, the core challenge in this paper centers around extracting the LoS power at each device (without knowing the source (voice) signal).
\new

While this problem is very hard in general, we find that this ``facing direction'' application presents unique opportunities.
Briefly, we observe that all the microphones at $N$ devices can be combined to approximate the source signal.
This (rough) source signal can offer (rough) estimates of the channels at each device.
Through an iterative algorithm that again combines these channels, we jointly refine the source signal and the channel, until convergence is reached.  
At this point, we extract the LoS channel for each device (from the estimated multi-tap channels).
We equalize these LoS estimates and correlate against a global human radiation pattern.
The strongest correlation determines the facing direction.
\new

We have implemented {\name} on \hl{$N=8$ devices}, each device built on a Raspberry Pi mounted with a 6-microphone circular array (an off-the-shelf hardware from SEED \cite{ReSpeaker}).
We scattered these devices in various configurations and our pre-processing protocol plays short chirps on each of them to mutually localize the device topology.
We performed experiments \hl{in $5$ different rooms with $9$ users} giving voice commands from different locations.
One of the devices serves as the leader, i.e., it collects the voice signal-information from all devices, estimates the facing direction, and returns the results to everyone else.
This process completes in real time, hence each device knows the user's facing direction and location before the voice command is complete.
\new

Overall, our experiments were performed over \hl{$3,524$ voice commands}.
In more than $54$\% of tests, the facing device is correctly identified.
When mistakes occur, {\name} \hl{mostly detects the angularly nearby device.}
\hl{This is still useful as a spatial filter, facilitating natural language processing (NLP) tools to mitigate the remaining disambiguity (using the full voice command).}
{\name} also localizes each user with a median of $37$ to $68$cm, depending on the number of devices in the room (more devices results in better accuracy).
Finally, {\name} does not rely on data-based training, hence exhibits stability across rooms, users, and different device configurations.
\new

In sum, our contributions may be summarized as follows:
\squishlist
\item We define the problem of facing direction in the context of multiple voice assistants.
For this context, we develop a diversity combining algorithm that iteratively estimates the LoS power from a multipath-polluted signal.
\item We address a range of practical concerns in building the full system.
This includes appropriately combining the LoS power with beamforming, localization, and the use of human radiation patterns. 
\hl{The results are encouraging but leaves room for future improvements.}
\squishend

The rest of the paper expands on these contributions, beginning with the problem formulation, followed by our ideas, failed attempts, and the actual design.
We end with evaluation and limitations of the current {\name} system.

\section{Problem Formulation}
\begin{figure}[hbt]
\centering
\includegraphics[width=3.2in]{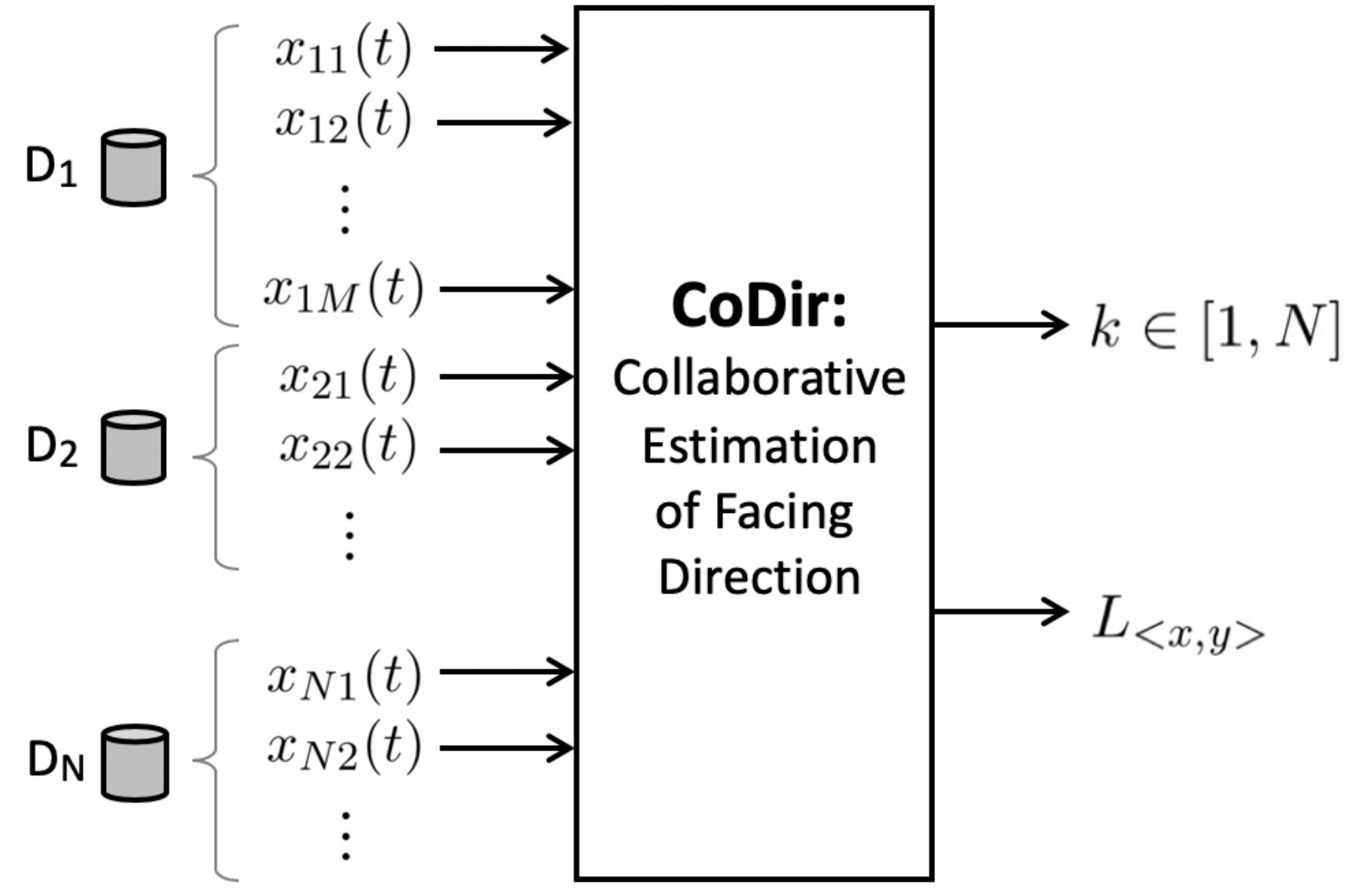}
\minus \minus
\caption{{\name} \hl{accepts up to} NxM microphone signals across all devices and outputs the facing device index ${k}$ and the user's location $L_{<x,y>}$.}
\label{fig:io} 
\vspace{-0.15in}
\end{figure}

Let us begin by formulating the ``facing direction'' problem. 
Let $D_{i},~i\in[1,N]$ denote each of the $N$ devices running voice assistants.
Each device has $M$ microphones, so let's denote the microphones as $m_{ij},~j\in[1,M]$.
Let's also denote the voice signal recorded by each microphone as $x_{ij}(t)$.
{\name}'s goal is to receive all the $x_{ij}(t)$ as inputs and output $k$, where $D_{k}$ is the device towards which the user is facing.
Figure \ref{fig:io} pictorially illustrates the problem definition.
In fact, as we discuss later, estimating $k$ will require that the user's location be computed, so {\name} will also output the user's 2D location $L_{<x,y>}$.
\new

$\blacksquare$ {\bf Assumptions:} We allow the following assumptions:
\squishlist
\item[1.] Each device has at least $M=4$ microphones to estimate angle of arrival (AoA) of voice signals.
\item[2.] The user faces the device when issuing the voice command, i.e., 
the face and upper chest is turned towards the device. 
\item[3.] The location and orientation of one of the devices is known with respect to the map of the house.
This is necessary to align the reference frames of the device-configuration and the house.
\squishend

Importantly, we do not assume the user's location, the user's voice radiation pattern, or any knowledge about the voice signal.  
We also do not assume time synchronization between the $N$ devices.
We expect {\name} to estimate the facing direction in sub-second granularity, i.e., before the user has completed the voice command.

\section{Basic Approach and Its Issues}
\begin{figure}[hbt]
\centering
\includegraphics[width=\columnwidth]{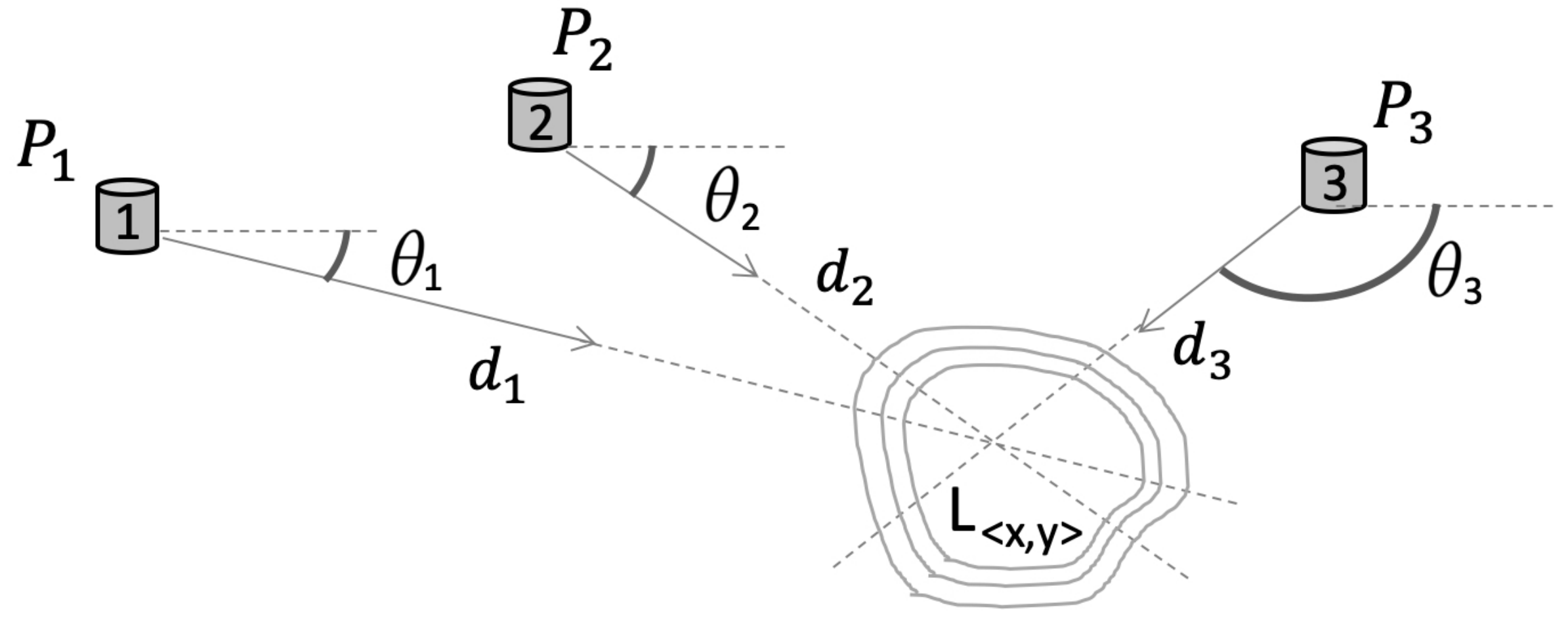}
\minus \minus \minus \minus
\caption{Basic overall operation with {\name}.}
\label{fig:basic} 
\vspace{-0.15in}
\end{figure}

Our high level approach can be expressed in $5$ simple steps.
Figure \ref{fig:basic} aids in following these steps.
\begin{enumerate} 
\item Each device $D_{i}$ computes the angle of arrival (AoA) $\theta_{i}$ of the voice signal and the power of the received signal, $P_{i}$.
\item The leader device gathers these information and first uses the $\theta_{i}$s to triangulate the user's location $L_{<x,y>}$.
The distance $d_{i}$ from the user to each device is now known.
\item The power $P_{i}$ at each device is now ``equalized'' by the distance $d_{i}$ according to the equation: 
\begin{displaymath}
{P}^{*}_{i} = P_{i} \times d_{i}^{2}
\end{displaymath}
The reason is to remove the effect of distance on received power, and let the equalized power across devices be only a function of the human radiation pattern. 
\item Pick the max among the $N$ equalized powers, say $P^{*}_{k}$.
Output $k$ and $L_{<x,y>}$.
\minus
\end{enumerate} 

\subsection{Issues}
\begin{figure}[hbt]
\centering
\includegraphics[width=3.2in]{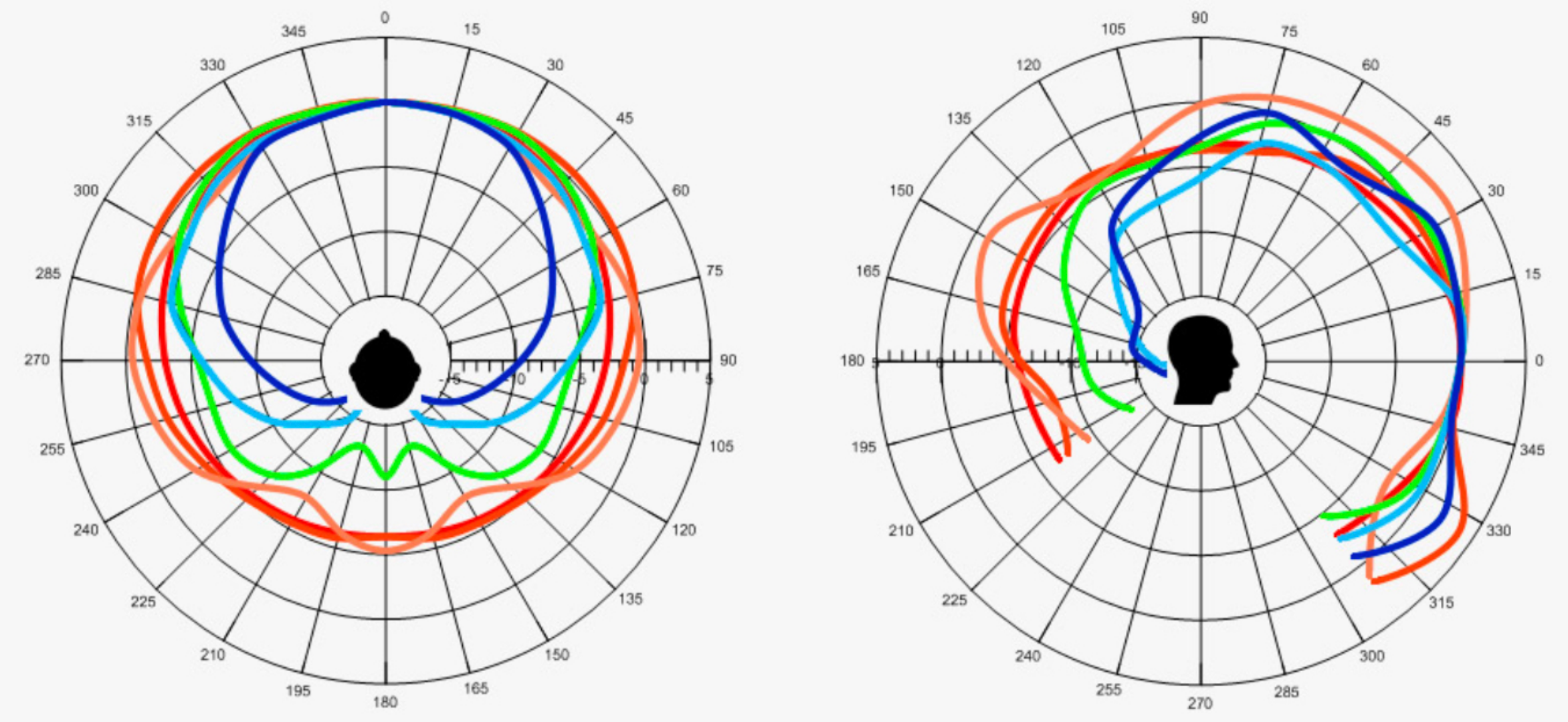}
\minus \minus
\caption{Example voice radiation pattern at discrete frequencies (a) horizontal (b) vertical cross section.}
\label{fig:radiation} 
\vspace{-0.05in}
\end{figure}

There are multiple hurdles in realizing this high level approach. 
First, AoA estimation in step-1 would incur errors in real indoor environments, translating to location error and ultimately errors in $d_{i}$. 
Second, the computed signal power $P_{i}$ is polluted by multipath and ambient noise.
The idea of equalization in step-3 is only meaningful if the power $P_{i}$ is the line-of-sight (LoS) power.
Thus, without the LoS power and with incorrect distance $d_{i}$, 
equalization will fail.  
This is particularly so because the voice radiation pattern's \hl{main lobe is not significantly stronger than the side/back lobes (see example in Figure} \ref{fig:radiation}).
This leaves a small margin for error. 

There is, however, some good news.
(1) Observe that $2$ AoAs are sufficient for localizing the user on the 2D plane.
When we have $N \geq 3$ devices, the redundant AoAs will overdetermine the system, reducing the localization and $d_{i}$ errors.
(2) The other good news is that the precise facing direction angle (say $\phi$) is not needed; we only need to select one facing direction from $N$ options, like a multiple choice question.
(3) Finally, instead of picking the max of $P_{i}^{*}$ in step-4, pattern matching will provide robustness, 
i.e., the shape of the voice radiation pattern can be matched against the shape of equalized powers across devices.
In sum, these application-specific opportunities make the problem more forgiving, offering the needed error margin to LoS power estimation.

\subsection{Key Technical Challenge}
Thus, the central challenge comes down to estimating the LoS power of the voice signal at each device.
These estimates of $P^{(LoS)}_{i}$ need not be accurate, but need to preserve the relative ordering between them.
The rest of the paper will largely focus on this specific challenge, followed by some engineering modules to pull the whole system together (namely, P2P device localization and AoA based user triangulation).
\minus

\subsection{Failed Attempts}
We faced a series of failures when extracting $P^{(LoS)}_{i}$ from $P_{i}$.
We briefly discuss them to motivate our final design 
\new

\textbf{Failure 1: Blind Channel Inference (BCI).} \\
The signal received at any microphone $x_{ij}$ is essentially the voice signal $v(t)$ arriving over the LoS path, plus delayed copies (i.e., echoes or multipath), arriving from different reflectors.
This can be modeled as $x_{ij} = v(t) * h_{ij}$, where $h_{ij}$ is the corresponding air channel.
In many applications (e.g., Wifi), part of the source signal is known to the receiver (called the preamble), hence $h_{ij}$ can be computed quite accurately.
The first tap of the $h_{ij}$ filter is the LoS path; knowing this is sufficient to estimate $P^{(LoS)}_{i}$.
\new

Unfortunately, the source signal $v(t)$ is unknown in our application because human voices vary widely even when the same user says the same word. 
Estimating the channel blindly (i.e., without knowing the source signal) has been extensively studied under blind channel inference (BCI) \cite{tong1994blind, muquet2002subspace}.
The core idea is to consider two nearby microphones, $m_{ij}$ and $m_{ik}$, and model their signals as 
\begin{displaymath}
x_{ij} = v * h_{ij}  \hspace{0.3in} x_{ik} = v * h_{ik} 
\end{displaymath}
BCI now attempts to search for $h_{ij}$ and $h_{ik}$ until
\begin{displaymath}
x_{ij} * h_{ik} = x_{ik} * h_{ij} 
\end{displaymath}
Since this search space can be very large, \hl{and because the objective function above is non-convex}, constraints are imposed on $h$, such as sparsity and ``similarity'' between $h_{ij}$ and $h_{ik}$. 
Under certain assumptions, the optimization converges, yielding the two channels and hence the LoS components.
\new

In real, indoor, multipath environments, BCI proved highly unstable.
Even when it converged, hours of computation was necessary on a multi-core machine. 
The reason for this is rooted in many factors, including hardware noise, ambient noise, and the fact that the channel is not as sparse as modeled in theory.
Given that we need the LoS component in sub-second granularity, we abandoned the BCI approach.
\new

\textbf{Failure 2: Diversity Combining.} \\
We considered the idea of combining the signals across all the microphones on the same device.
The hope is that the LoS components from $M$ microphones can add up coherently, while the multipath and noise components will be incoherent, resulting in a LoS dominated signal.
To this end, we applied a matched filter (i.e., delay-and-summing the microphone signals for every possible angle of arrival, and choosing the maximum).
We now have $N$ such combined signals -- one per device -- and checked if their maximum corresponded to the facing device (or at least if the $N$ signals displayed the shape of a voice radiation pattern). 
Results were inconsistent again.
The multipath components were often correlated due to the proximity of the microphones; moreover, with $M=4$ or $6$, there was not enough statistical diversity.
\new

\begin{figure}[hbt]
\centering
\includegraphics[width=\columnwidth, height=1.5in]{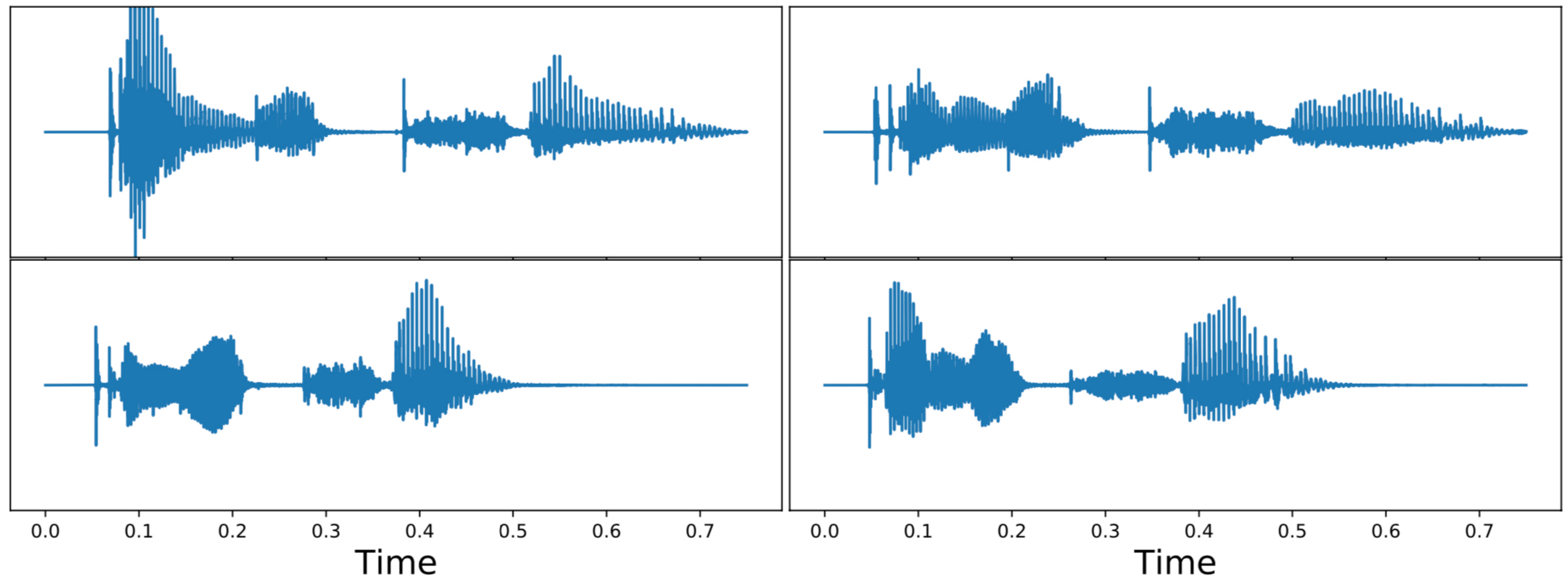}
\vspace{-0.25in}
\caption{Same word repeated by a user (recorded very close to the user's mouth) show significant variation.}
\label{fig:voice} 
\vspace{-0.15in}
\end{figure}

\textbf{Failure 3 and 4: Opportunities from Voice Signals.} \\
We looked for new opportunities that are specific to human voice/speech signals.
One opportunity was that the first few samples of the received signal (i.e., when the user just starts speaking) is only the LoS path; the echoes arrive a few milliseconds later.
We expected the power of these initial samples to be dominated by the LoS.
However, the results were again unreliable, and analysis showed that \hl{human voices ramp up slowly (Figure} \ref{fig:voice}).
Hence, the initial samples have low SNR (and are dominated by noise), and by the time the voice signal ramps up above the noise floor, the multipath components have arrived and polluted the signal.
The effect was pronounced for people with relatively softer voices, hence the idea was not generalizable.
\new

In a bolder attempt, we leveraged the information that a voice command would start with a wake-word, like ``Alexa'' or ``Ok, Google''.
We synthesized a voice signal for this wake-word (produced by a text-to-speech tool), and attempted to estimate the channel using this (rough) source signal.
We even asked a specific user to say the wake-word in various different ways, and used these templates for the source signal.
Unfortunately, the channel estimates were sensitive to variations of the source signal, and more importantly, the variations of the human voice were far larger than we hoped for (Figure \ref{fig:voice}).
We discarded this idea as well.

\section{The {\name} Algorithm}
This section presents our proposed algorithm for extracting $P^{(LoS)}_{i}$ from the received signal $x_{ij}$.
We focus on the core technique, assuming the other modules (like device and user localization) are solved.
The next section will discuss those modules and pull together the end to end {\name} system.

\subsection{Intuition}
Let us begin with a high level intuition -- this will not explain the algorithm but should show the line of thinking. 
Our aim is to harness the maximum amount of diversity gain possible, and therefore use all $N\times M$ microphones together.
Although the powers $P^{(LoS)}_{i}$ are very different across the $N$ devices, the source voice signal is the same in all of them.
If we can reasonably estimate the source signal (i.e., estimated $\hat{v}(t)$ is in the neighborhood of the true voice signal $v(t)$), 
then it might be possible to estimate channels $\hat{h}_{ij}$ that are also in the neighborhood of the true channel $h_{ij}$.
Now, if we diversity-combine these channels (i.e., {\em the LoS components are coherently aligned while multipath/noise add incoherently}), we expect to estimate a better source signal (closer to $v(t)$).
This can continue iteratively, with updated channels improving the source signal estimate, in turn improving the channel estimates.
Once converged, the first channel tap in each of $\hat{h}_{ij}$ should correspond to the LoS component, which we hope will lead us to $P^{(LoS)}_{i}$.
Let us now explain the actual algorithm. 

\subsection{Primitives}
\begin{figure}[t]
\centering
\includegraphics[width=3.2in]{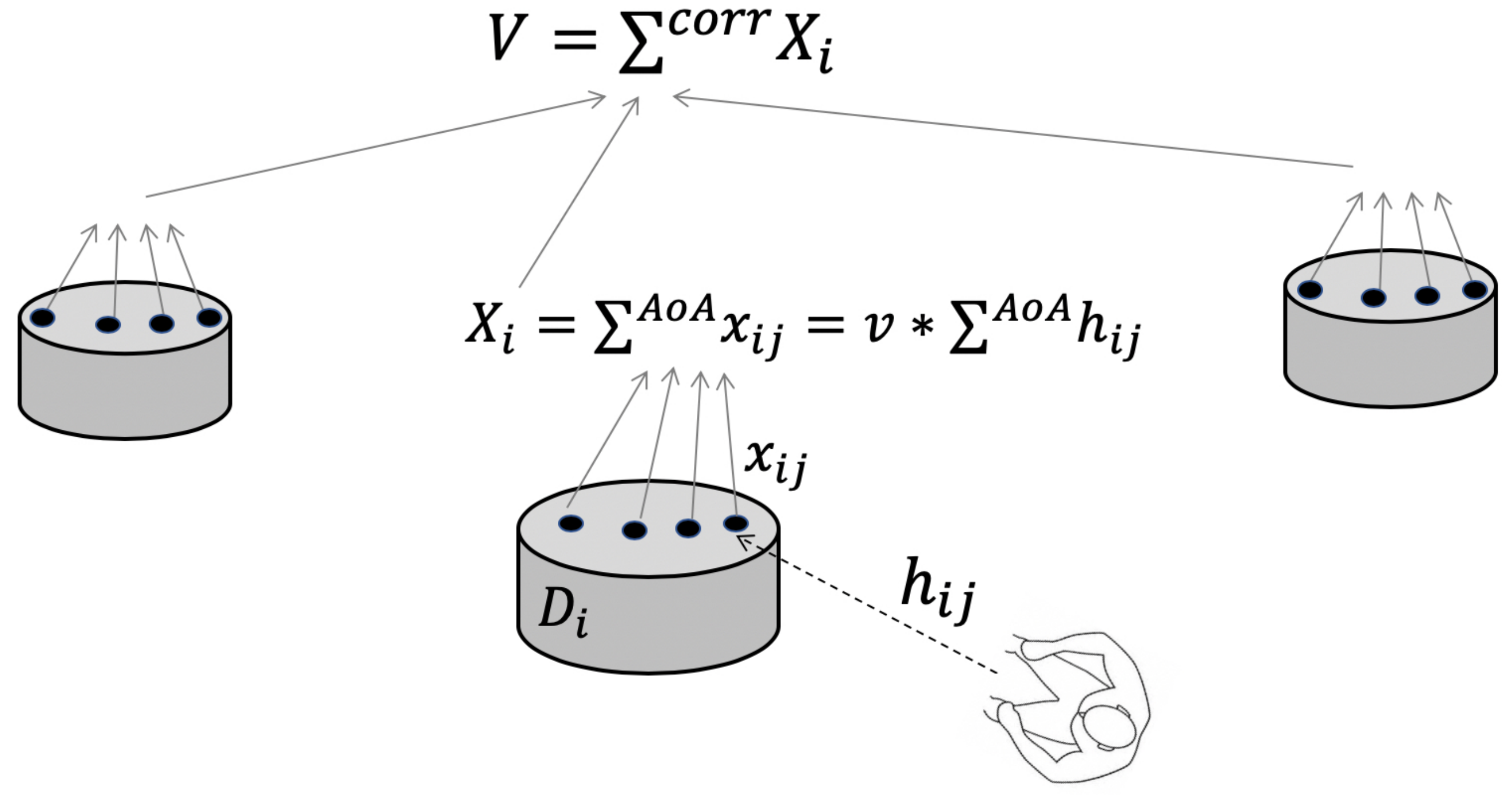}
\minus \minus
\caption{\hl{Two signal combination primitives: Locally within a device, and Globally across devices.}}
\label{fig:primitives} 
\vspace{-0.05in}
\end{figure}

We first define two signal combination primitives -- Local and Global -- as visualized in Figure \ref{fig:primitives}.
\new

\textbf{(1) Local Delay and Sum on AoA, denoted as $\sum^{AoA}$}:  \\
At a given device, we align the signals from $M$ local microphones based on the signal's AoA. 
This is a standard technique that {\em coherently} aligns the strongest signal component (LoS), while multipath and noise add {\em incoherently}.
We denote this locally combined signal as $X_{i}$ and is expressed as:
\begin{align*}
    X_{i} = \sum\nolimits^{AoA}_j x_{ij}(t) & \coloneqq \sum_j x_{ij}(t-\Delta_j)
\end{align*}
where $\Delta_j$ is the relative delay at microphone $m_{ij}$ from a reference microphone $m_{i1}$.
Of course, $\Delta_j$ is a function of the AoA $\theta_i$, but since the user and device locations are estimated by {\name} (discussed in Section \ref{sec:design}), $\theta_i$ is known here.
\new

\textbf{(2) Global Align on Correlation, denoted as $\sum^{Corr}$}: \\
We now combine signals from different devices, $D_{i}$.
AoA alignment is not possible anymore since the signals across the $N$ devices have no meaningful delay relationships.
Hence, we aim to align the strongest signal component from each device.  
For this, we correlate and estimate each device's relative delay; each signal $X_{i}$ is then delayed accordingly and summed up to \hl{form $\hat{V}$}.
We define 
\begin{align*}
   \hat{V} = \sum\nolimits^{Corr}_j X_j(t)\coloneqq \sum_j X_{j}(t-\delta_j)
\end{align*}
where $\delta_j = {argmax}_{\delta} (Corr(X_0(t),X_j(t-\delta)))$.
\new


\subsection{The Iterative Algorithm}
Algorithm \ref{alg:fd} presents the pseudo code for {\name}, so let's zoom into the details.
There are $6$ main steps. 
\new

\textbf{Step 1: Each device locally aligns microphone signals}: 
This results in a per-device signal $X_{i} = \sum^{AoA}_j x_{ij}$, which can be written as:
\begin{align}
    X_{i} &= \sum\nolimits^{AoA}_j v * h_{ij} = v * \sum\nolimits^{AoA}_j h_{ij} = v * h_i
    \label{eq:localhi}
\end{align}
This $h_{i}$ channel slightly suppress ambient noises and multipath (slightly because ambient noise and multipath are quite correlated across closely located microphones on the same device).
However, thermal noise gets considerably mitigated in this step since they are independent/uncorrelated.
\minus

\makeatletter
\newcommand{\skipNo}{%
      \let\old@ALG@lno=\ALG@step%
        \renewcommand{\ALG@step} {%
            \global\let\ALG@step=\old@ALG@lno}%
            }
\makeatother
\begin{algorithm}[htb]
\caption{\hl{Facing Direction} $(x_{ij}, RadiationPattern)$ }
    \begin{algorithmic}[1]
\State $X_{i} = DelaySumAoA (x_{i1}, x_{i2} ... x_{iM})$
\State $\hat{v}^{0} = AlignCorrelate (X_{1}, X_{2} ... X_{N})$

\skipNo
\State $j=0$
\skipNo
\While {True}
\State $h_{i}^{j} = Deconvolve (X_{i}, \hat{v}^{j})$
\skipNo
\State $h = PeakAlign (h_{1}^{j,+}, h_{2}^{j,+} ... h_{N}^{j,+})$
\State $\hat{v}^{j+1} = Deconvolve (\hat{v}^{j}, h)$

\skipNo
\State $j=j+1$
\skipNo
\If{$|\hat{v}^{j}-\hat{v}^{j-1}|<\Delta$}
\skipNo
\State break
\EndIf
\EndWhile
\State $A_i= FirstPeak(h_{i}^{j})$
\skipNo
\State $P^*_i=A_i^2\times d_i^2$
\State $k = argmax_k Corr(P^*, Radiation Pattern)$
\end{algorithmic}
\label{alg:fd}
\minus
\end{algorithm}

\textbf{Step 2: Construct approximation of source signal $\hat{V}^0$}: 
The $X_{i}$ signal from each device is forwarded to the leader and globally combined.
This achieves better multipath suppression and is a first estimate of the source signal, say $\hat{V}^0(t)$.
\begin{align}
    \hat{V}^0 &= \sum\nolimits^{corr}_i X_i &= \sum\nolimits^{corr}_i v*h_i &= v*\sum\nolimits^{corr}_i h_i
    \label{eq:est_v}
\end{align}


Now, consider $\sum^{corr}_i h_i$.
This is a single global channel, say $h$, in which the strongest taps from all $h_{i}$ have added constructively, while the other (multipath) taps are much weaker since the channels are uncorrelated across devices.
Accordingly, we have
\begin{align}
    h \coloneqq \sum\nolimits^{corr}_i h_i(t) &= \sum_i h_i(t-argmax(h_i))
    \label{eq:hglo}
        \minus
\end{align}
And the strongest tap of this channel $h$ is:
\begin{align*}
    max(h) &= \sum_i max(h_i) 
        \minus
\end{align*}
which is $\approx N$ times larger than any $h_i$.
This {\em sharpened} channel -- with the maximal tap much higher than others -- is of high SNR and makes the aligned signal $\hat{V}^0$ a good approximation of the source voice signal $v$.
\new

\textbf{Step 3: Use $\hat{V}^0$ to update local/global channels:} \\
Pretending $\hat{V}^{0}$ to be the source voice signal (i.e., $X_{i} = \hat{V}^0 * h_{i}$), we first estimate local channels $h_{i}$ via deconvolution: 
\begin{align*}
\hat{h}_i^0=deconvolve(X_i,\hat{V}^0)
\minus
\end{align*}
Then, we update the global channel $h$ as
\begin{align}
\hat{h}^0=\sum^{corr}_i \hat{h}_i^{0,+}
\label{eq:h0}
\minus
\end{align}
where $\hat{h}_i^{0+}$ only preserves the positive values of $\hat{h}_i^{0}$ and zero-forces negative amplitude taps (i.e., we are boosting the SNR of the LoS path)\footnote{
From a mathematical perspective, removing negative parts of $\hat{h}_i$ makes the function non-linear, 
otherwise we would get $\hat{h}$ as the delta function, i.e.,
\begin{align*}
    \sum\nolimits^{corr}_i \hat{h}_i &= \sum\nolimits^{corr} deconvolve(X_i, \hat{V}^0)\\
    &= \sum\nolimits^{corr}_i deconvolve(v*h_i, v*\sum\nolimits^{corr}_i h_i)\\
    &= deconvolve(\sum\nolimits^{corr}_i v*h_i, v*\sum\nolimits^{corr}_i h_i)\\
    &= \delta(t)
\minus
\end{align*}
}.
\new

\textbf{Step 4: Iteratively converge on voice signal $v$ from $\hat{V}^j$}: \\
From Equation ~\ref{eq:est_v} and \ref{eq:hglo}, we know that the estimated source signal $\hat{V}^{0}$ 
is actually the true voice signal $v$ convolved with the global channel $h$. 
Now, Equation~\ref{eq:h0} gives us the estimate for $h$. 
By plugging this back into Equation~\ref{eq:est_v}, we can estimate a better source signal.
Hence, we deconvolve $\hat{V}^0$ with $\hat{h}^0$ to get $\hat{V}^1$.
Pretending $\hat{V}^1$ as the new/better source signal, we start the next iteration from \textbf{Step 3} (where $\hat{V}^0$ is updated to $\hat{V}^1$).
The iterations continue towards convergence, ultimately yielding channel $\hat{h}_{i}$ for each device.
\new

\textbf{Step 5: Extract LoS power from local channels $\hat{h}_{i}$}: \\
We stop the iterations when $h_{i}$ converges, i.e., when the difference of estimated channels between consecutive iterations is smaller than a threshold $\Delta$. 
At this point, we take the estimated channel $\hat{h}_i$ of device $D_i$ and record the amplitude of the first peak $A_i$.
This amplitude is used \hl{to calculate the LoS power for} $D_{i}$ as: 
\begin{align}
P_i^{LoS} = A_i^2
\minus
\end{align}
We then equalize by the distance as $P_i^* = P_i^{LoS}\times d_i^2$.
\new

\begin{figure}[hbt]
\minus \minus
\centering
\includegraphics[width=0.48\columnwidth]{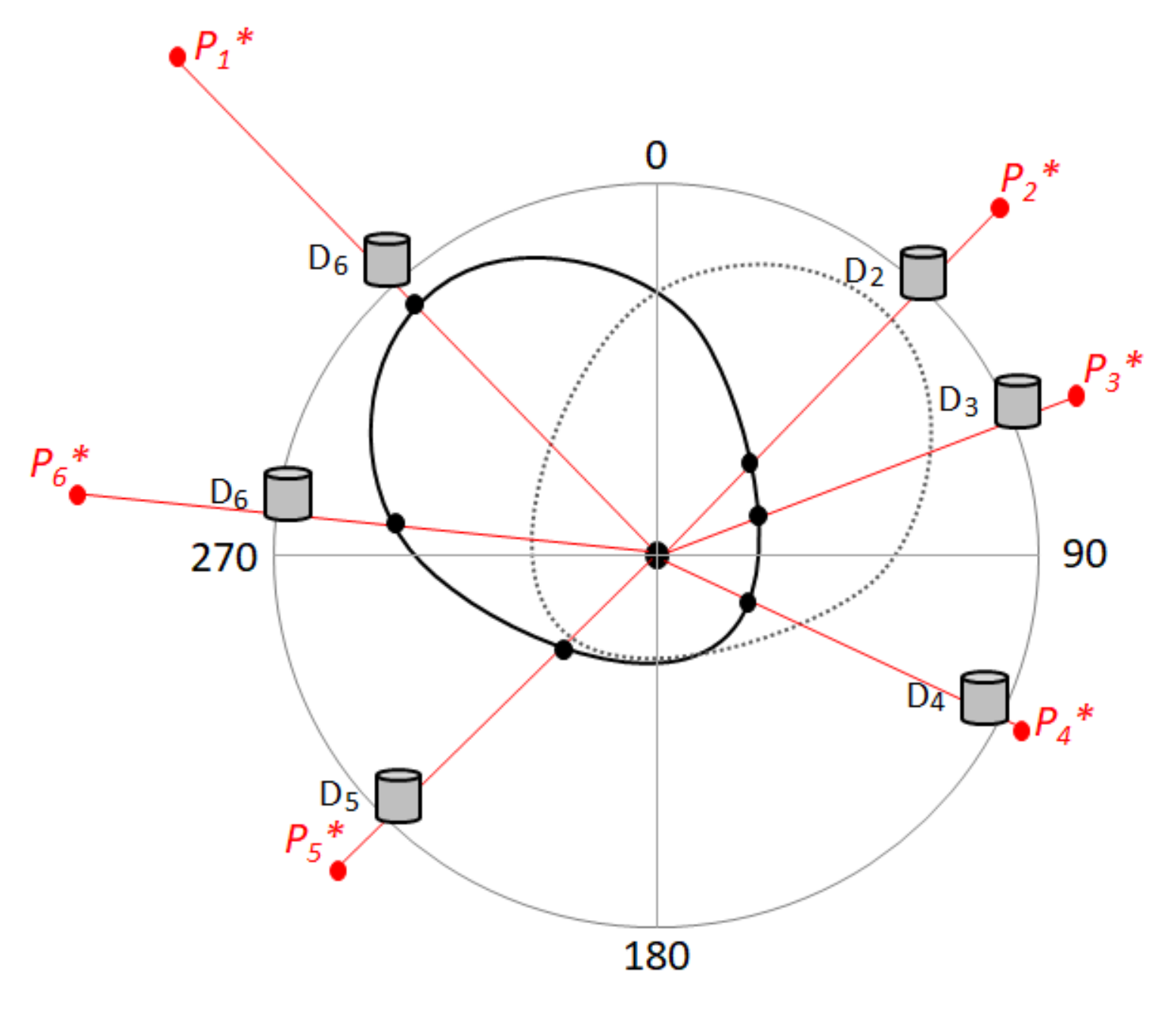}  
\hfill
\includegraphics[width=0.48\columnwidth, height=1.3in]{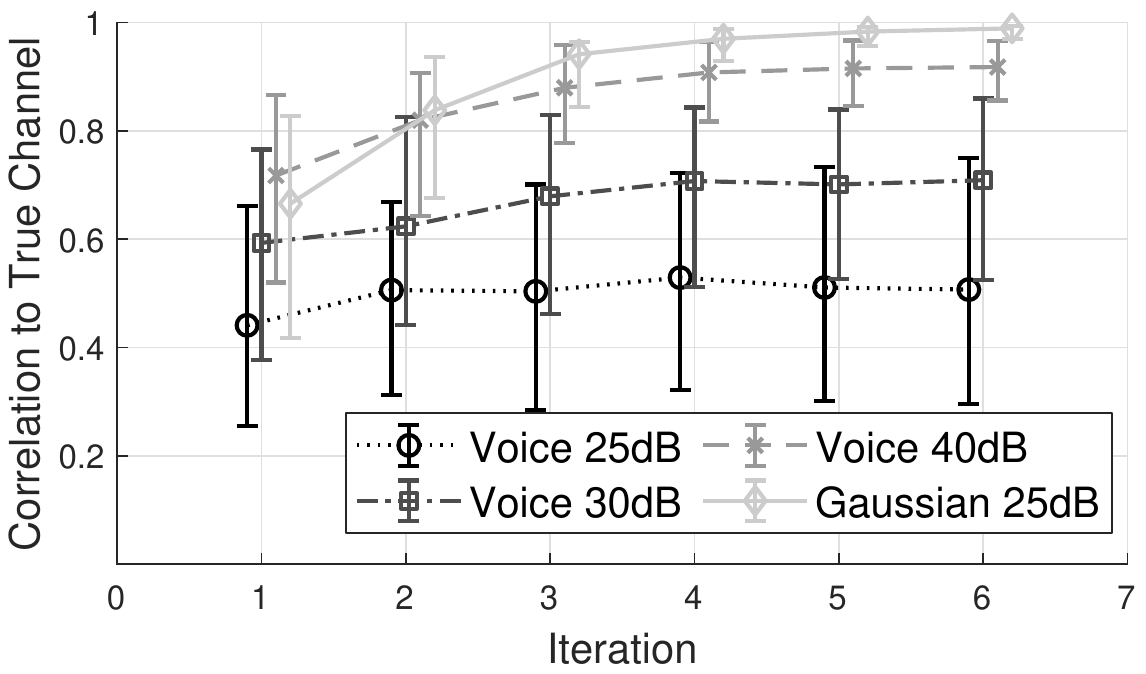}
\minus \minus
\caption{(a) Correlating equalized powers from devices (red dots) with voice radiation pattern (black dots). (b) Simulation of convergence of algorithm.}
\label{fig:correlate} 
\vspace{-0.15in}
\end{figure}

\textbf{Step 6: Correlate Equalized Power with radiation pattern}: 
Correlate the $N$ equalized powers with the voice radiation pattern (note this is a circular correlation as shown in Figure \ref{fig:correlate}(a)).
Identify the facing device $D_k$ with the max correlation, and report the index $k$.
We observe that speech radiation patterns are similar across different users, and show in Section~\ref{sec:eval} that using a universal radiation pattern does not affect the accuracy.

\subsection{Convergence Simulations}
We do not have a proof of convergence for the $h_{i}$ estimation.
However, we simulated a wide range of multipath scenarios -- using true speech waveforms as the source signal -- and matched the estimated $\hat{h}_{i}$ with the correct/known $h_{i}$.
Our matching metric takes the first tap from $N$ estimated channels, and cross-correlates this vector with the true first taps of the same $N$ devices.
Figure \ref{fig:correlate}(b) plots the cross-correlation across iterations on the X axis.
\hl{Evidently, the convergence is encouraging with SNRs at $30$dB and higher (as is the case with voice assistant).
When the source signal has less auto-correlation (e.g., a random Gaussian signal), results are even better.}
We will visit more evaluations later; for now, we turn to the rest of the {\name} system, including P2P localization and user triangulation.

\section{System Design}
\label{sec:design}

\begin{figure*}
\includegraphics[width=\linewidth]{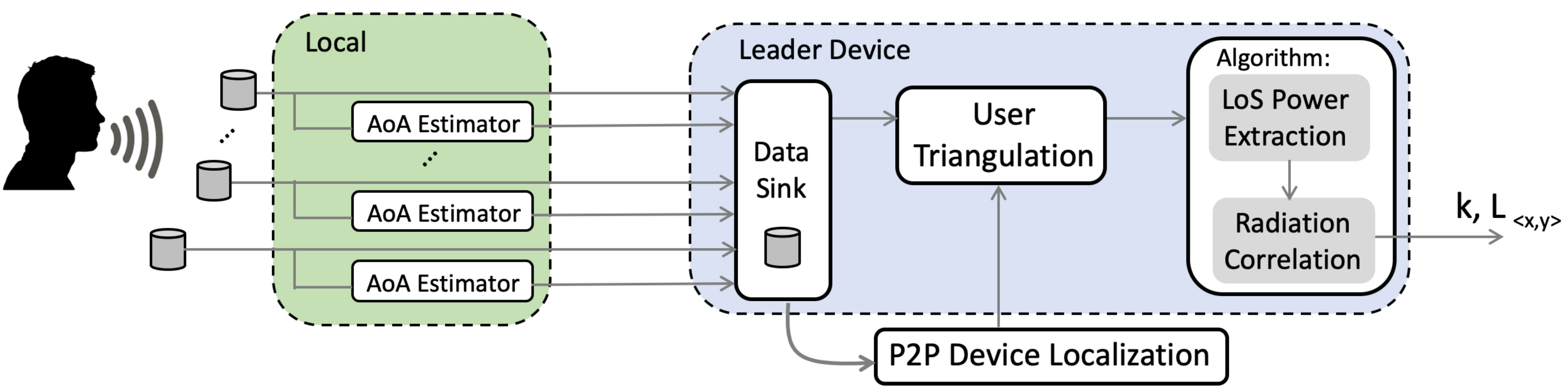}
\minus \minus \minus \minus \minus
\caption{System architecture of CoDIR: Each device runs the AoA Estimator locally and forwards the AoA and the voice signals to the leader device (cloud, edge, or a leader device). The leader performs signal processing across all the signals, inferring the user location and facing direction.}
\vspace{-0.1in}
\label{fig:system}
\end{figure*}

Figure \ref{fig:system} shows the end-to-end system pipeline.
Each device $D_i$ activates on recognizing a wake-word like ``Alexa'', and runs an \texttt{\ul{AoA Estimator}} locally to derive the signal's direction $\theta_{i}$. 
The recorded signals and the AoAs are forwarded to a leader device -- the cloud, an edge router, or just one of the voice assistants. 
The leader uses the AoAs to first triangulate the user's 2D location.
Since \texttt{\ul{User Triangulation}} requires the device locations, the leader runs \texttt{\ul{P2P Device Localization}} during setup to infer their spatial configuration.
Together, the result of user triangulation yields the user's distance to each of the devices -- these distances and the raw signals are then forwarded to the {\name} algorithm.
As described in the previous section, the algorithm outputs the target device $k$ and user location $L_{<x,y>}$.

\subsection{P2P Device Localization}
During installation, and periodically after that (perhaps every night or when some devices have moved) the {\name} system runs a device localization protocol.
The electronic speaker embedded in each voice assistant broadcasts a short chirp; other devices overhear the chirp and compute their own AoAs using the AoA Estimator as discussed below.
\new

\textbf{AoA Estimator.}
AoA estimation is performed locally for parallel operations (i.e., saves time at the leader device).
Since only the LoS angle is necessary, we adopt the well-established generalized cross-correlation phase transform (GCC-PHAT) algorithm \cite{knapp1976gcc}. 
GCC-PHAT essentially computes the correlation between two signals \hl{but normalizes with the product of their amplitudes to avoid amplitude biases}.
The max correlation value $G$ is recorded per microphone-pair, which leads to the AoA estimate.
\new

Standard GCC-PHAT implementations operate on a single microphone pair that subtends the largest acute angle to the direction of the signal. 
We modify this to compute the AoA for $\frac{M}{2}$ pairs, i.e., the diagonally opposite ones. 
To be specific, we take signals from each microphone pair $p \in [1,\frac{M}{2}]$ of device $D_i$ and compute cross-correlation values $g^p_i$ for that pair. 
Since microphones on a given device are located very close to each other, GCC-PHAT does give correlations at a fine AoA granularity. 
Hence, we interpolate the correlation values $g^p_i$ at a granularity of $1^{\circ}$, and average them across all microphone pairs, denoted as $\hat{g}^p_i$. 
The final AoA=$\theta_i$ is obtained by taking the position of the peak of the cross-correlation values $\hat{g}_{i}^p$ as follows: 
\begin{align}
\theta_i = argmax_{\theta} |\hat{g}^p_i|
\label{eq:gcc1}
\end{align}

{\bf P2P Localization:} 
Note that the chirp signal is globally known, so each device need not compute AoA {\em blindly}.
Instead, a device $D_{i}$ extracts the channel for the known chirp (via deconvolution), and uses the channel-pairs as the input to GCC-PHAT.
Denoting an estimated channel pair as $h_{ij}$ and $h_{ik}$, the AoA is estimated as:
\begin{displaymath}
\theta_{i,jk}=gccphat(h_{ij}, h_{ik})
\end{displaymath}
Upon interpolating and averaging over the microphone pairs, the final $\theta_{i}$ is forwarded to the leader device.
Note that devices take turns in broadcasting the chirp, so the AoAs are computed for each broadcast.
These redundant measurements help over-determine the system.
\new

Let us denote $\theta_{i}^{j}$ as the AoA at device $D_{i}$ from $D_{j}$.
The leader first removes all unreliable $\theta_{i}^{j}$, i.e., those $\theta_{i}^{j}$ that are far from $\theta_{j}^{i}+180^\circ$. 
This is possible since the leader knows all the AoAs.
Finally, the locations $L^{D_i}_{<x,y>}$ of devices are optimized as:
\begin{displaymath}
argmin_{\mathbf{x, y}} \sum_{i,j} |angle(L_{<x,y>}^i - L_{<x,y>}^j)-\theta_{i}^{j}|
\end{displaymath}
Essentially, this optimization computes locations of all devices that best satisfies the P2P AoA constraints available to the leader.
\new

It is worth noting that the estimated device locations are in the reference frame of the devices (not an absolute location).  
However, if this reference frame is aligned with the floorplan of the place, the device and user locations can become meaningful to many other apps (e.g., the user is in the kitchen or the living room).
\hl{One possible way to achieve this is to ask the user to stand at two known locations on the floorplan, speak the wake-word, and have the devices localize them.}
Once this is done, the device and floorplan's reference frames are aligned, offering semantics to the user's behavior.
\minus \minus

\subsection{User Triangulation}
Assuming the system is set up (i.e., device configuration is known), a user now speaks the wake-word ``Alexa''.
Observe that GCC-PHAT must now operate {\em blindly} \hl{on the raw signal (Equation} \ref{eq:gcc1}) and estimates the AoA.
The leader receives these AoA which may be viewed as rays emanating from each device towards the human speaker.
Ideally, the rays should intersect at a single location -- the user location. 
In reality, AoA and P2P location errors cause the rays to deviate, producing many intersection points.
To cope with outliers, we borrow a clustering scheme, DBScan \cite{ester1996density}. 
\new

DBScan is a non-parametric density-based clustering algorithm that groups closely packed points and eliminates low-density groups.
We configure DBScan with a minimum cluster size of $2$ points and a maximum cluster diameter as $1$m.
The cluster with the highest number of points is retained and its centroid is declared as the user's location $L_{<x,y>}$. 
{\name} outputs this location with its computed facing device, $k$.
\minus

\section{Evaluation}
\label{sec:eval}

\subsection{Implementation}
Our ``devices'' are implemented on a Raspberry Pi4s ($64$ GB) \cite{RaspberryPi}, mounted with a $6$-microphone circular array from SEEED ReSpeaker \cite{ReSpeaker}.
Figure~\ref{fig:respeaker}(a) shows a complete device.
The arrangement of the microphones are similar to Amazon Echo or Google Home, but unlike these devices, the raw acoustic samples are available from the SEEED at $16$kHz.
A native speech recognition system detects wake-words like ``Alexa'' or ``Snowboy'' and forwards the signal segment to {\name}.
The {\name} modules run on the Pi, and depending on whether a device is a follower or leader, data is transmitted/received.
Finally, the leader device computes the facing direction and notifies all the devices -- a green LED lights up on only the facing device.
\new

For P2P device localization -- where devices need to play short sounds -- we place a small XiaoMi Bluetooth speaker~\cite{XiaoMi} below each device.
The sounds and data exchanges are coordinated by the leader device over WiFi (using socket connections).
At run time, a leader device is chosen arbitrarily since any device can take on this role. 
Our overall framework is implemented in Python and is extensible to variations, such as BLE, Zigbee, or offloading to edge devices.

                    
\subsection{Experiment Methodology}
\begin{figure*}[hbt]
\includegraphics[width=0.565\linewidth,trim= 0cm 0.5cm 0cm 1cm,clip=true]{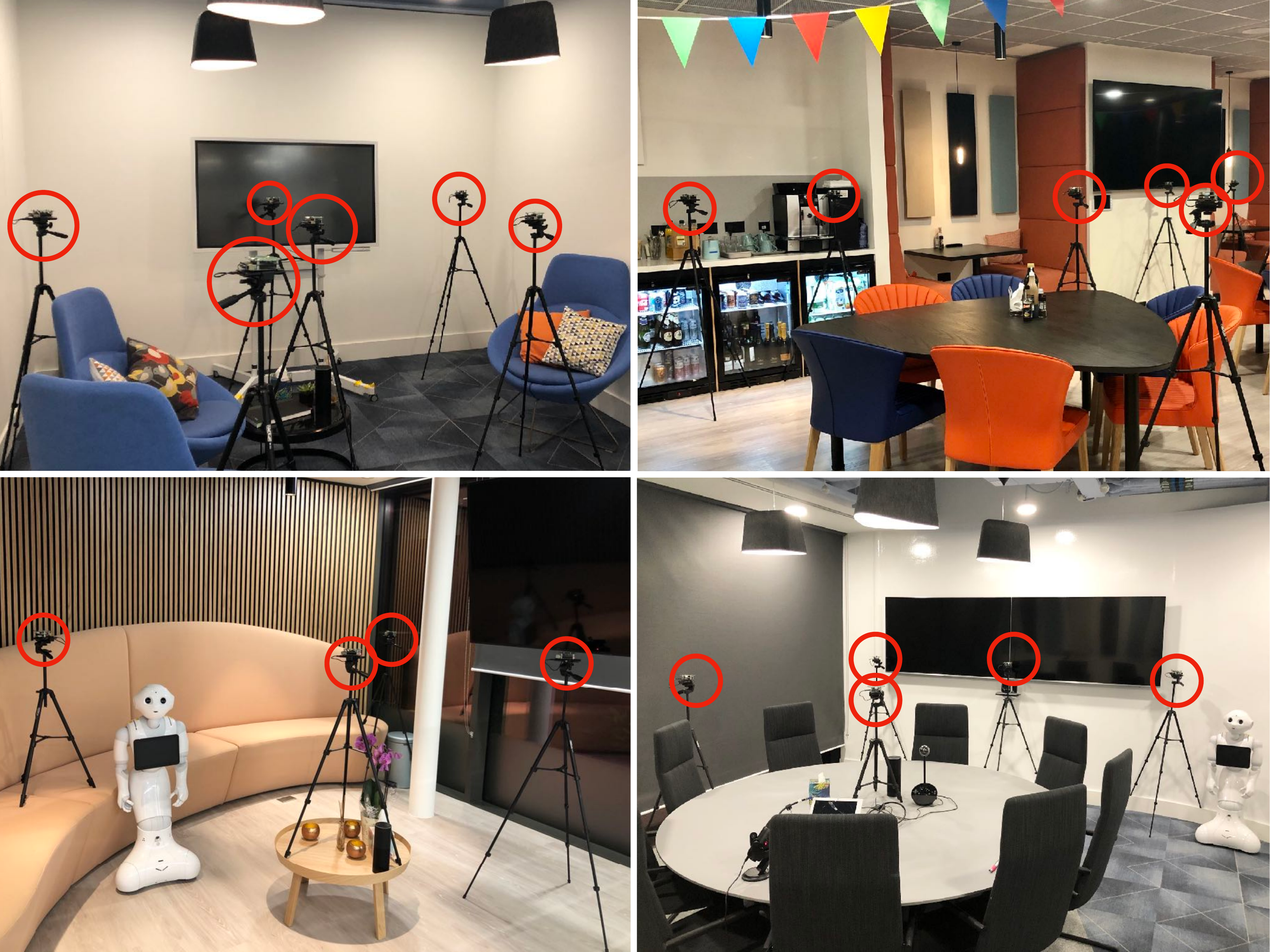}
\includegraphics[width=0.43\linewidth]{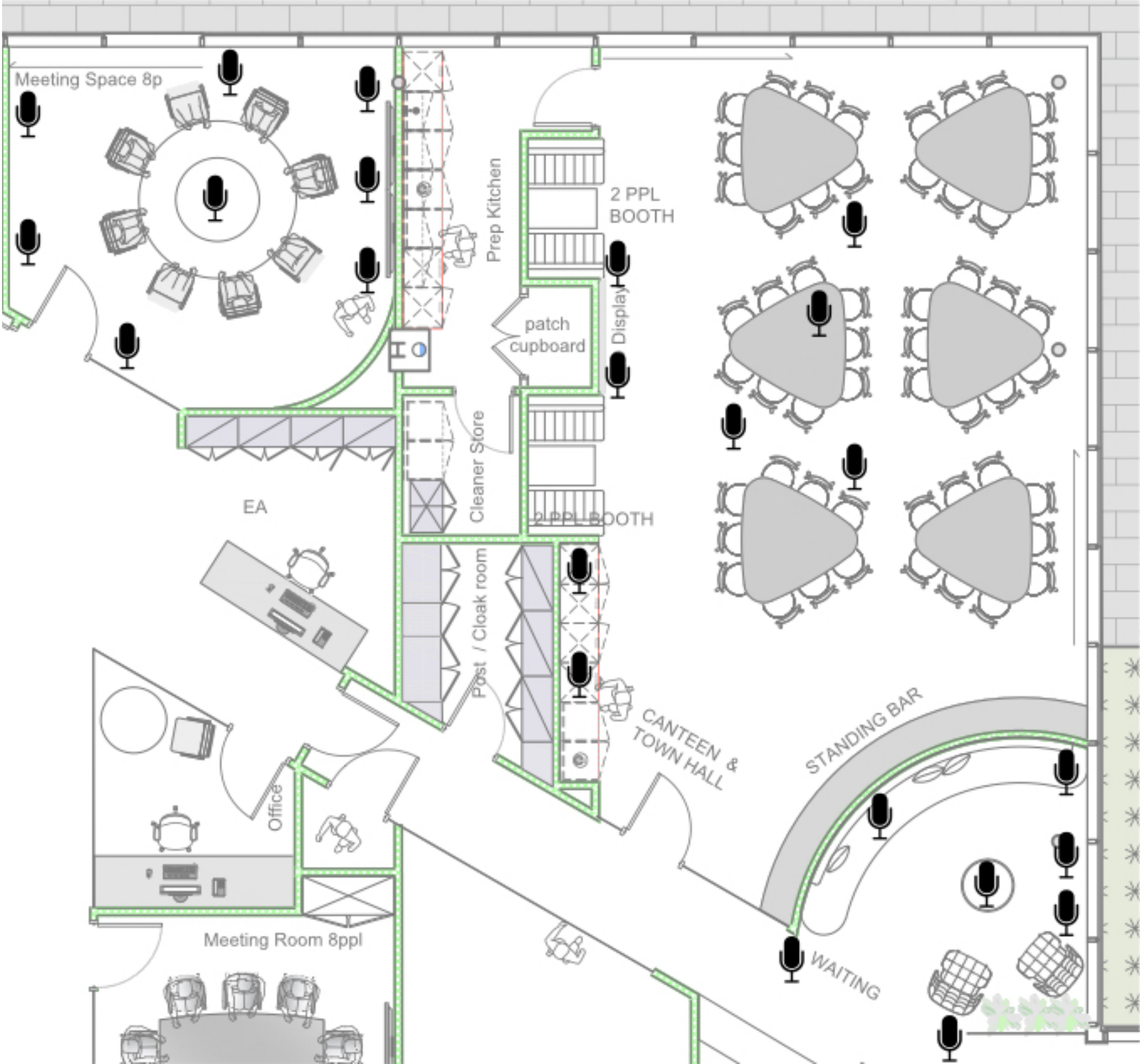}
\caption{Four different rooms and a floor plan of the rooms with microphone icons indicating device placements. These environments are multipath rich, as expected in real houses and offices.}
\minus \minus
\label{fig:rooms}
\end{figure*}

Our experiments are executed with $N=8$ devices, scattered in hundreds of configurations \hl{across $5$ different rooms.}
We often place the devices next to existing gadgets (such as TVs, microwaves, thermostats, faucets), pretending that they are embedded inside.
Figure~\ref{fig:rooms} shows $4$ rooms and a floorplan of the building, indicating that these are not controlled environments (instead are rich in multipath as would be the case in real homes and offices).
For voice commands, we \hl{recruit $9$ adult volunteers}, almost equally divided between males and females, and ask them to speak while standing, seated, or even walking slowly.
Ground truth is measured using a laser ranger, so the locations of devices and volunteers are known to a millimeter resolution.
\new

During experimentation, we invite each volunteer to pick arbitrary locations in a room and speak the wake-word $8$ times, facing each device. 
From this data, we pick all subsets of $2$, $4$, and $6$ devices, and evaluate facing direction accuracy for these configurations.
This covers a wide range of scenarios -- some in which all devices are on one side of the user, while others in which the user is surrounded by devices.
These experiments are repeated over multiple configurations, rooms, and users.
\hl{Altogether, we record $>15,000$ audio clips resulting in $>20,000$ facing direction tests.}

\subsection{Success Metric}
\begin{figure}[hbt]
\centering
\includegraphics[width=0.48\columnwidth]{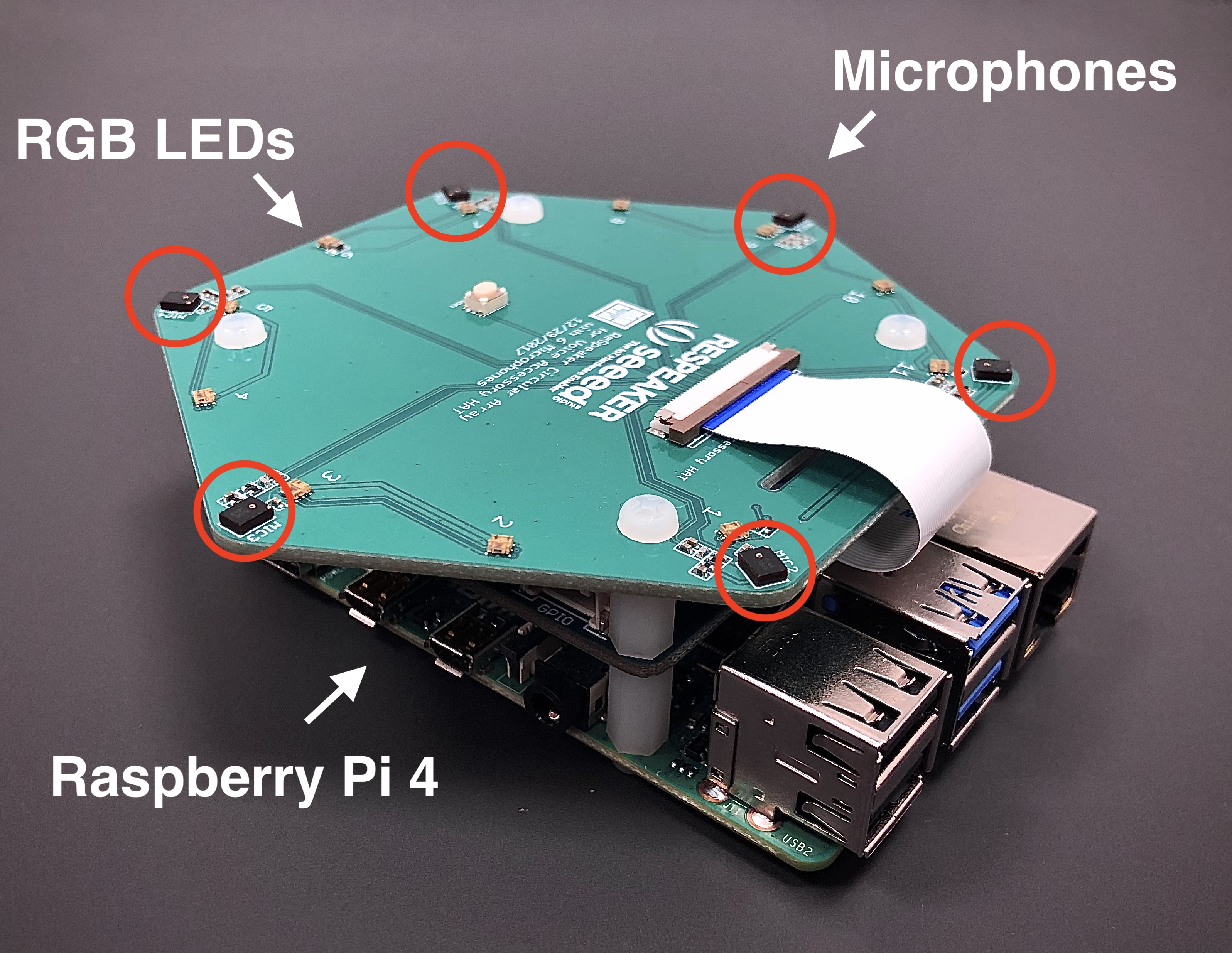}
\includegraphics[width=0.48\columnwidth, height=1.25in, trim= 0cm 0cm 0cm 0.8em,clip=true]{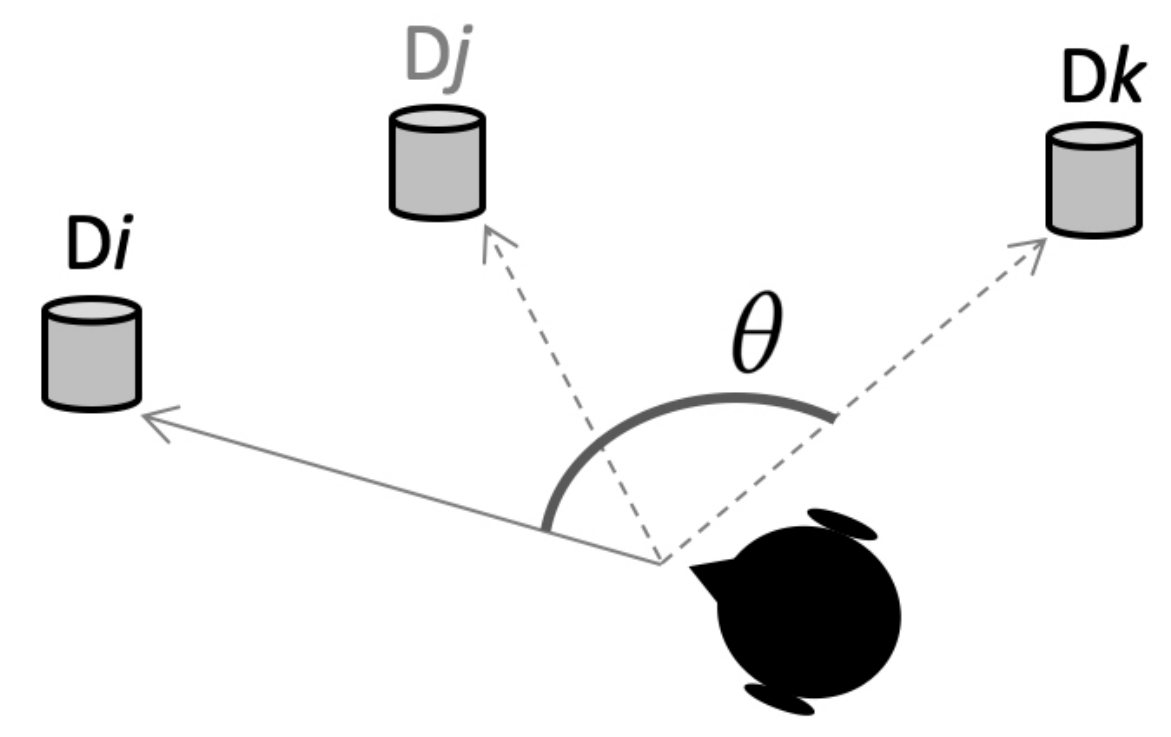}
\caption{(a) A device composed of 6-Mic circular array mounted on Raspberry Pi4. (b) Facing direction error (FDE) and facing integer error (FIE) to characterize the nature of errors.}
\vspace{-0.15in}
\label{fig:respeaker}
\end{figure}

We define two main metrics of evaluation, namely {\em Facing Direction Error (FDE)} and {\em Facing Integer Error (FIE)}.
Consider Figure \ref{fig:respeaker}(b) where the user is truly facing $D_{i}$, and assume that {\name} determines $D_{k}$ as the facing device. 
In this case, $FDE=\theta$ and $FIE=2$ since $D_{k}$ is angularly separated by $2$ devices from $D_{i}$.
Now, if $D_{j}$ was not present, then FDE is still $\theta$ but FIE would now be $1$.
This captures that {\name} could not have done any better than $\theta$ since there is no other device less than $\theta$ degrees away from $D_{i}$

Of course, in addition to FDE and FIE, we will also measure user location errors, as well as other microbenchmarks such as variation across SNR, radiation patterns, number of microphones, anchors, user mobility, etc.
\new

    
\subsection{Performance Analysis}
\begin{figure}[hbt] 
\vspace{0.05in}
\includegraphics[width=\columnwidth, trim={1em 0 3em 3em},clip]{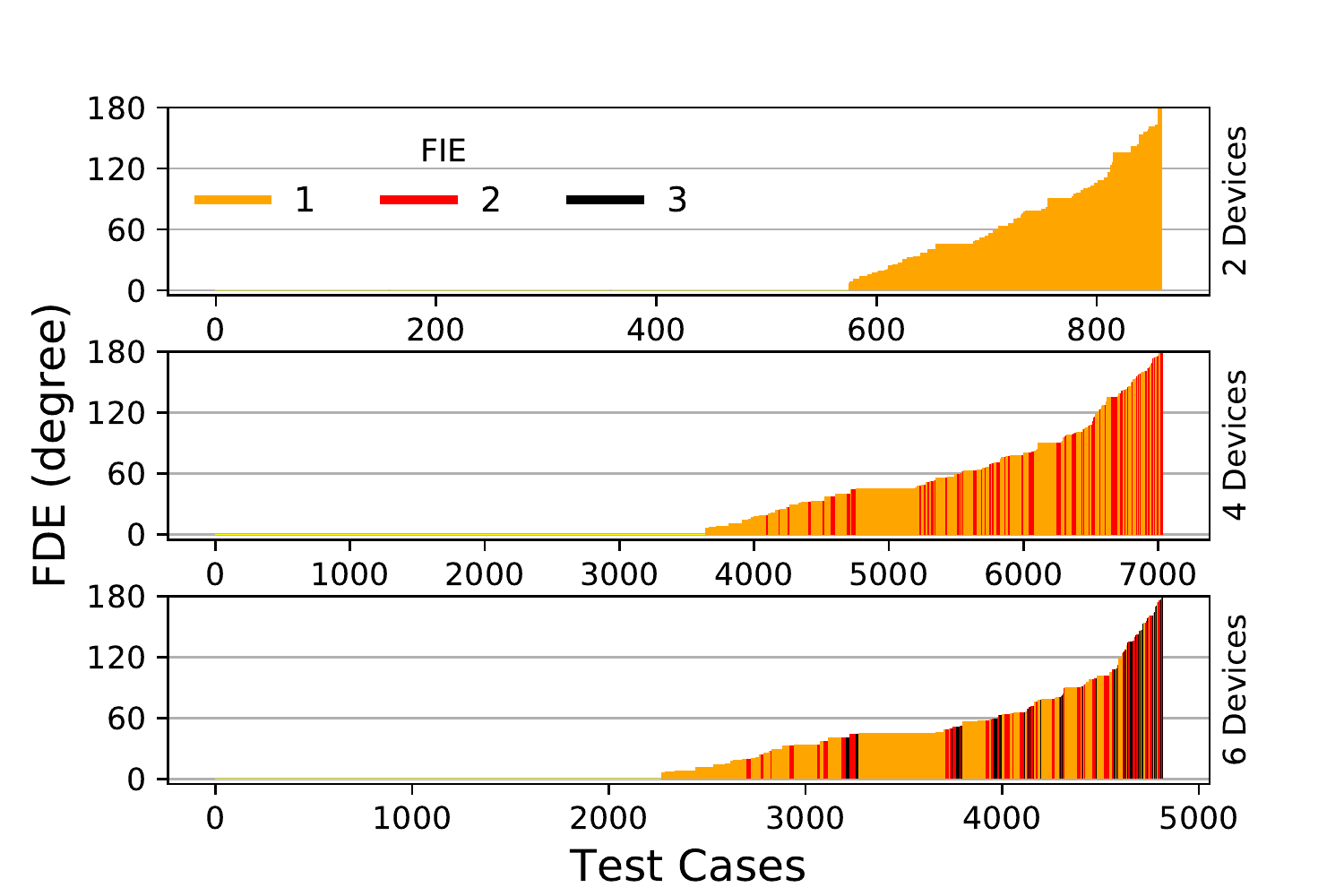}
	\vspace*{-2em}
            \caption{Facing direction error in all test cases employing 2, 4 and 6 devices. For each sub-figure, FDE is arranged in increasing order from left to right. We use 3 different colors to represent FIE (the number of missed devices). The darker the color, the greater is the number of devices between the ground 
truth and detected device.}
\vspace{-0.15in}
           \label{fig:fd_dt_loc}        
        \end{figure}

\begin{figure}[hbt] 
\vspace{0.05in}
\includegraphics[width=\columnwidth, trim={1em 0 3em 3em},clip]{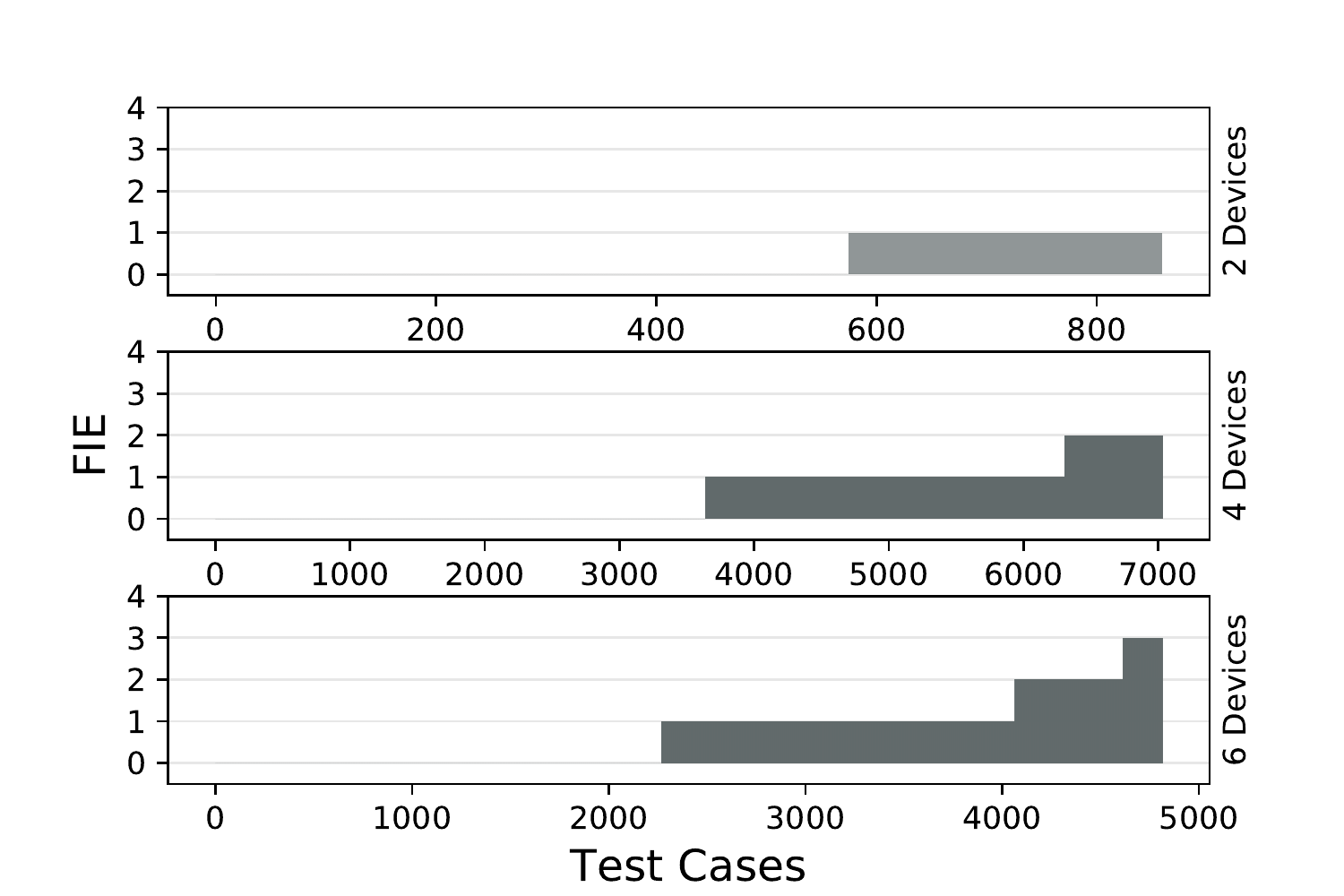}
\vspace*{-2em}
\caption{FIE across all test samples.}
\vspace{-0.15in}
\label{fig:fie}        
\end{figure}

$\blacksquare$ \textbf{Facing Direction Error}: 
Figure~\ref{fig:fd_dt_loc} shows the FDE and FIE errors for $2$, $4$, and $6$ devices.
The errors are derived from all the test configurations and sorted in increasing order.
The height of the bars indicate the FDE while their color indicates FIE.
For instance, with $4$ devices, the facing device is correctly identified in $52$\% of the test cases.
When errors occur, $79$\% of them exhibit an FIE of $1$, meaning the adjacent device is selected.
Figure \ref{fig:fie} shows the break up of FIE across all the test cases. 
Importantly, observe that 2-device scenarios are not much easier/better than $6$ because the user localization is better with $6$ devices, which aids in FDE/FIE.
Results confirm this as well, showing a similar error distribution across different device densities.
\new

\begin{figure}[h]
            \centering
              \vspace{-0.05in}
             \includegraphics[width=\columnwidth, height=1.6in]{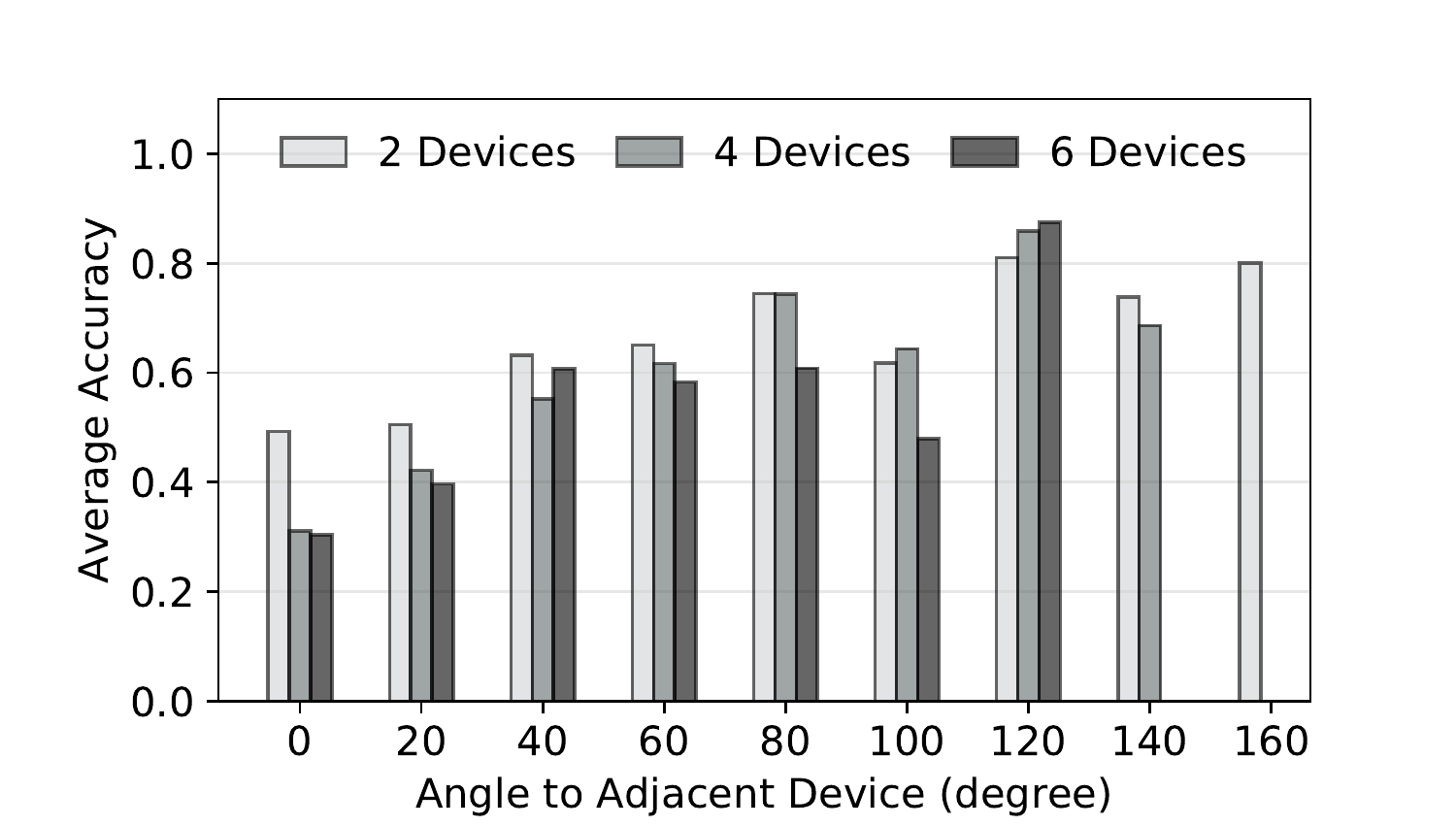}
	    \vspace{-0.2in}
            \caption{Estimation accuracy under different device separations, i.e. angle between the device user is facing
            and its adjacent device w.r.t the user.
            }
	    \vspace{-0.1in}
            \label{fig:sep2acc}
        \end{figure}

Figure \ref{fig:sep2acc} breaks down the errors in angular buckets to visualize the nature of mistakes in {\name}.
The X axis represents angular ranges, e.g., the value $X=0$ means that the angle between the facing device and its adjacent device is between $0-20^{\circ}$.
Similarly, $X=20$ indicates that the adjacent device is between $20-40^{\circ}$ from the facing device.
The bars show the fraction of tests (across all such scenarios) in which the facing device was accurately determined.
The accuracy improves generally with increasing angular separation.
This is intuitive because greater angular separation implies that the facing device has a proportionally stronger power compared to others.
This tends to improve the correlation with the voice radiation pattern.
\new

Of course, this may not always be the case because large separation near the target devices results in angular aggregation of the remaining devices, which causes the correlation to get biased.
On the other hand, when there is only less than $10^{\circ}$ between two devices, small user movements can affect accuracy.
After combining all these factors, {\name} achieves $50$, $35$, and $33$\% accuracy even when the adjacent device is less than $40^{\circ}$ apart from the target device.
\new

\hl{A natural question is: in the face of errors, which device should respond to the voice command?
One possibility is to delay the decision until the full command has been heard, and use the facing direction and command semantics jointly, to determine the intended device.}
For instance, if the facing device is either the TV or the adjacent floor light, and the voice command is ``Alexa, turn off in 10 minutes'', then the device that is currently ON can accept the command (another light that is ON but located in the opposite direction need not respond to this command).
\new

\begin{figure}[h]
        \centering
        \includegraphics[width=0.8\columnwidth, height=1.4in,  trim={2em, 0, 3em, 3em}]{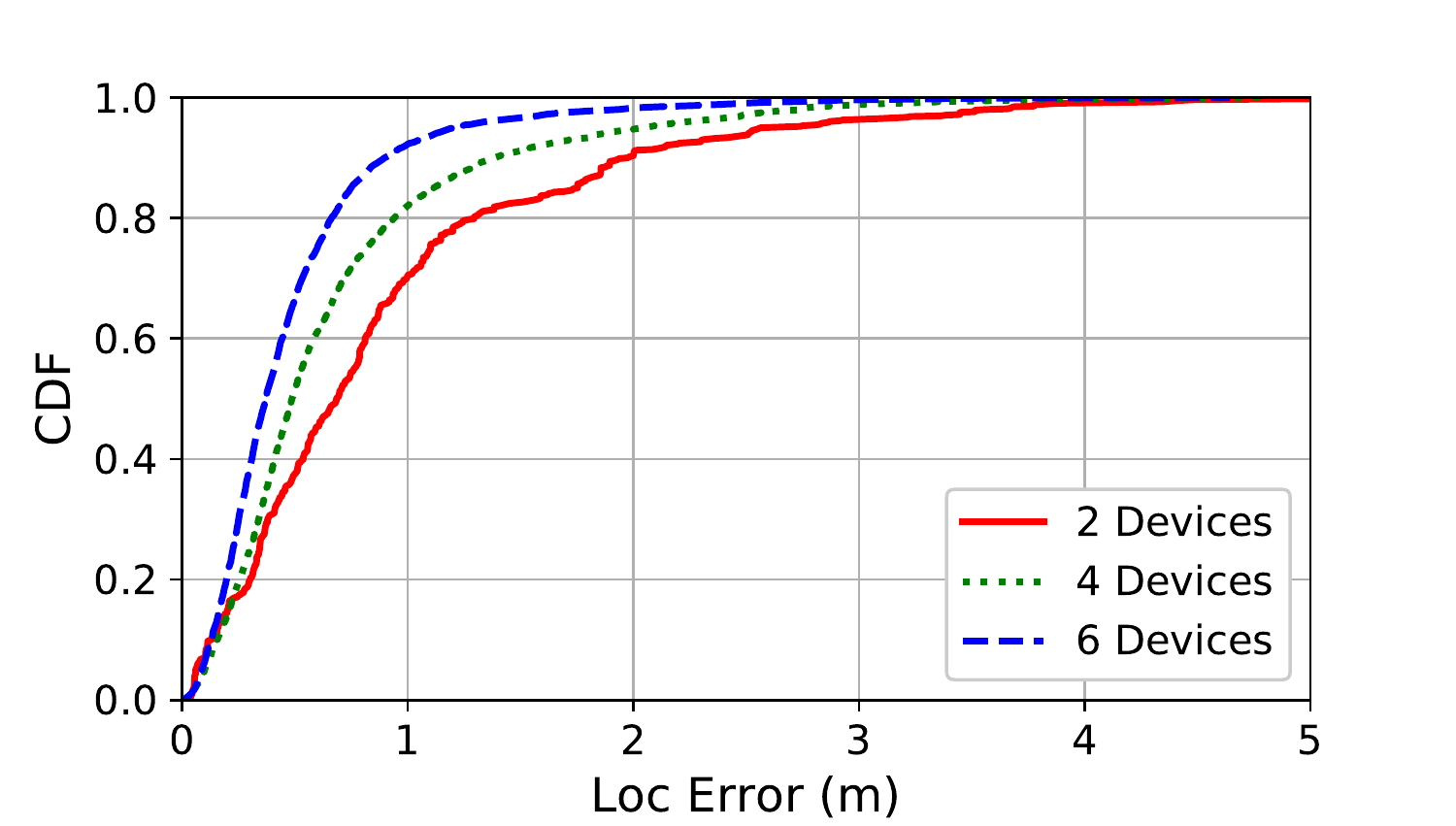}
	\vspace*{-1.2em}
        \caption{Localization error in all tested cases, employing 2, 4, and 6 number of devices.}
	\vspace*{-1.2em}
        \label{fig:loc_all}
    \end{figure}

$\blacksquare$ \textbf{User Localization Error}: 
Figure ~\ref{fig:loc_all} plots the user localization error as an outcome of AoA estimation and triangulation.
Results are again plotted for the $N=2$, $4$, and $6$ device settings.
The median errors are $0.7$m, $0.5$m, and $0.4$m, respectively.
The errors with $2$ devices are naturally higher since small AoA errors shift the intersection points during triangulation. 
With more devices, the system gets overdetermined, improving location accuracy.
With $6$ devices, for instance, around $90$\% of the test cases incur less than $1$m error.
Besides aiding in the facing direction problem, this user location can also be valuable to other applications that need indoor localization.
        
\subsection{Micro Benchmarks}

\begin{figure}[hbt]
            \centering
            \includegraphics[width=\columnwidth]{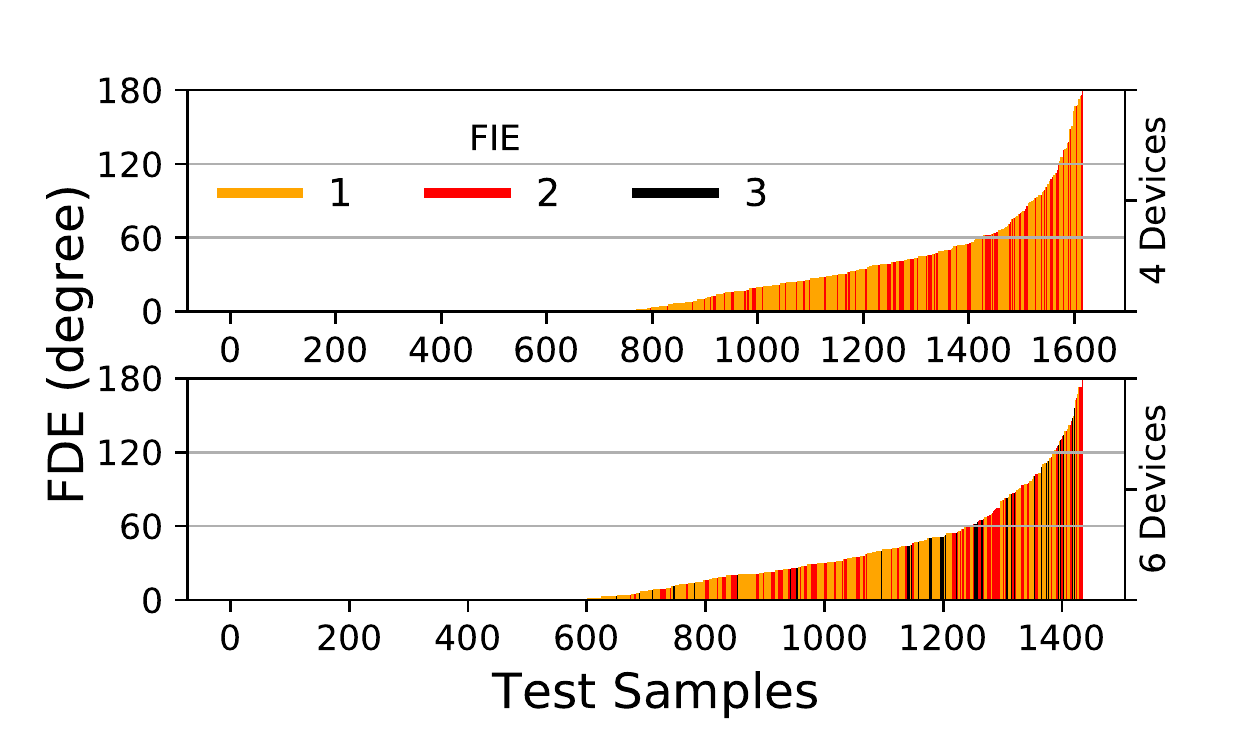} \vspace{-0.1in}
            \includegraphics[width=\columnwidth, height=1.8in]{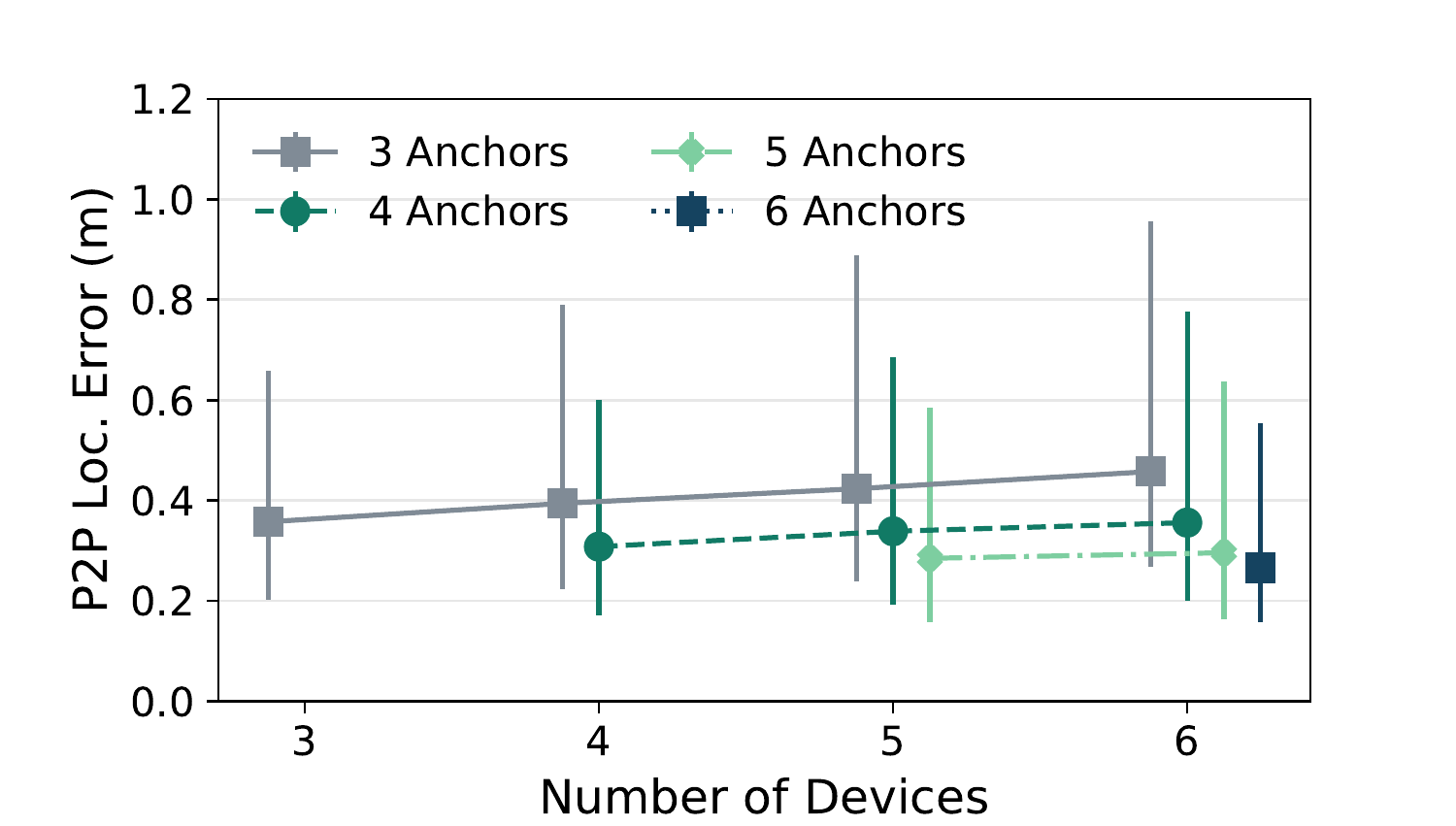}
	    \vspace*{-0.2in}
            \caption{(a) FDE when P2P localization is incorporated in {\name}. (b) Device localization error when using different $N$ and different subsets of anchor nodes.}
	    \vspace*{-0.1in}
            \label{fig:p2p}
            \end{figure}

$\blacksquare$ \textbf{Incorporating P2P:} 
Figure~\ref{fig:p2p}(a) illustrates the performance of {\name} when P2P device localization module is incorporated in the experiments.
Note that with 2 devices it is impossible to obtain the location of both devices, hence we show only results from $4$ and $6$ devices. 
Evidently, the P2P module does not degrade performance too much, implying the whole {\name} system can be made functional.
\new

    Figure ~\ref{fig:p2p}(b) shows the device location errors from the offline P2P localization module. 
    Since the P2P algorithm only reports the relative configuration of the devices,
    to calculate the positioning error, we scale the P2P estimation 
    using every device as the origin, calculate the distance to the ground-truth, and pick the minimum. 
    The choice of device (to serve as the origin) does not affect the facing direction algorithm, 
    but helps visualize the P2P accuracy. 

    \begin{figure*}        
        \begin{minipage}{\textwidth}
            \includegraphics[width=.33\linewidth,  trim={1em, 0, 3em, 2.5em}, clip]{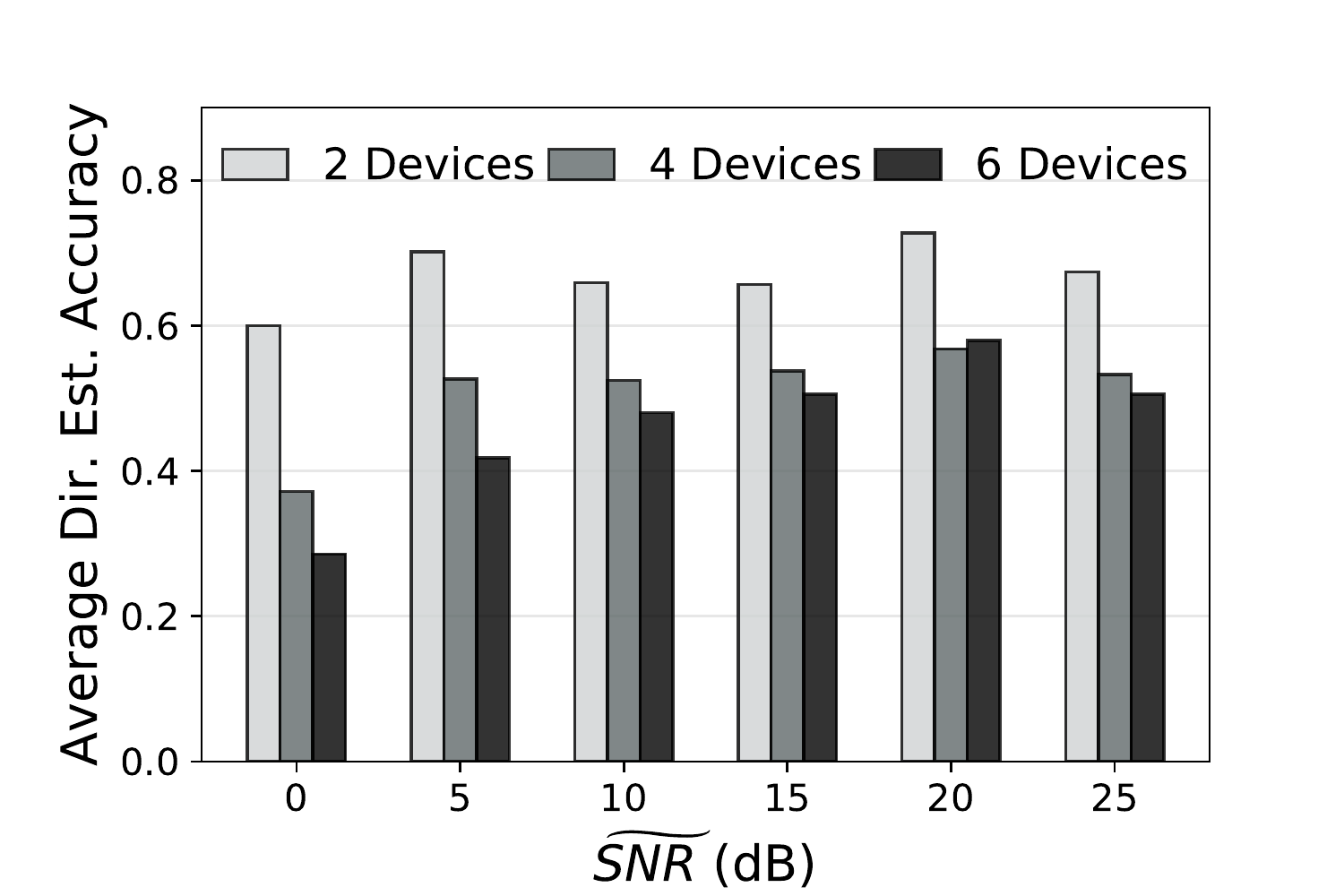}
             \includegraphics[width=.33\linewidth,  trim={1em, 0, 3em, 2.5em}, clip]{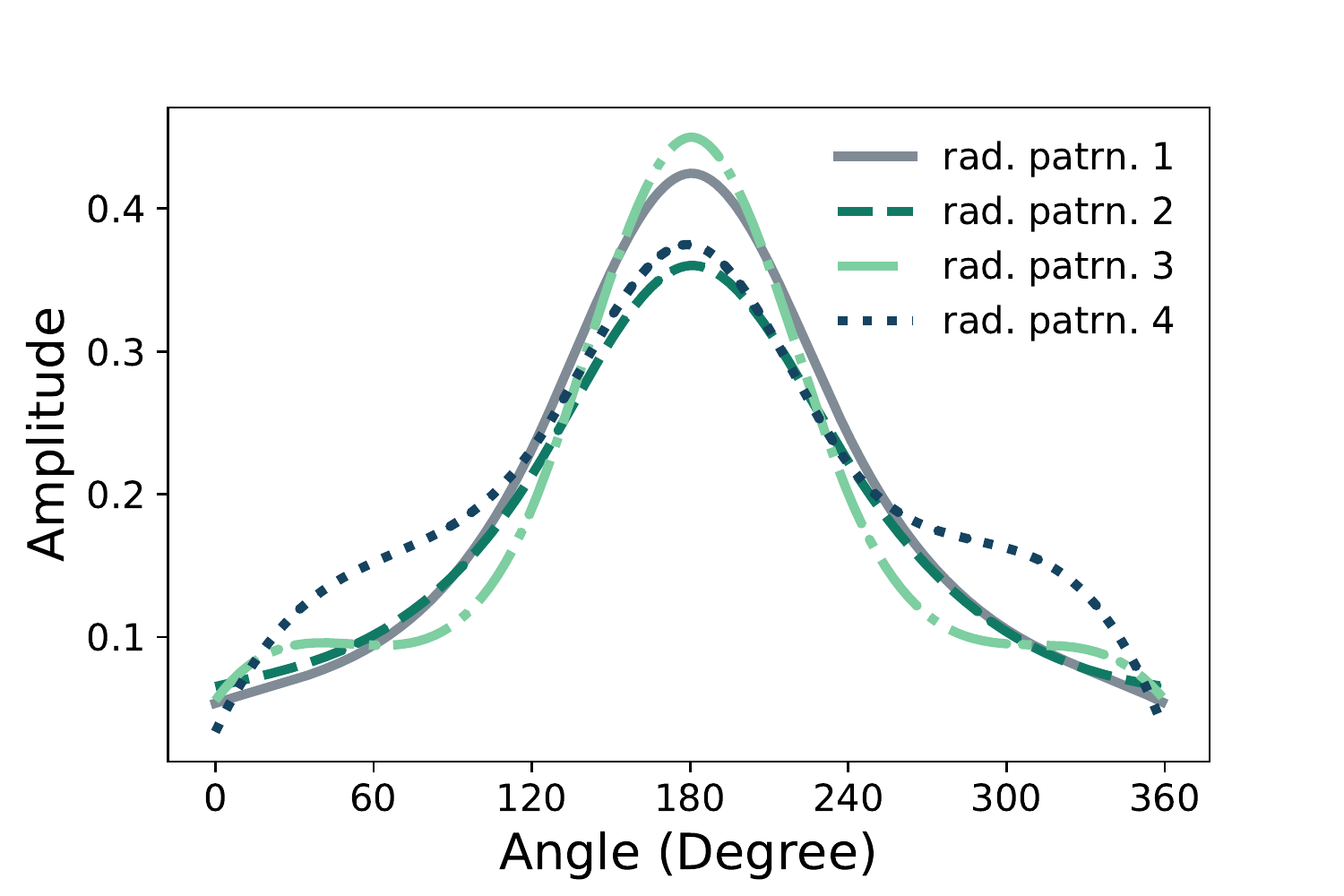}
             \includegraphics[width=.33\linewidth,  trim={1em, 0, 3em, 2.5em}, clip]{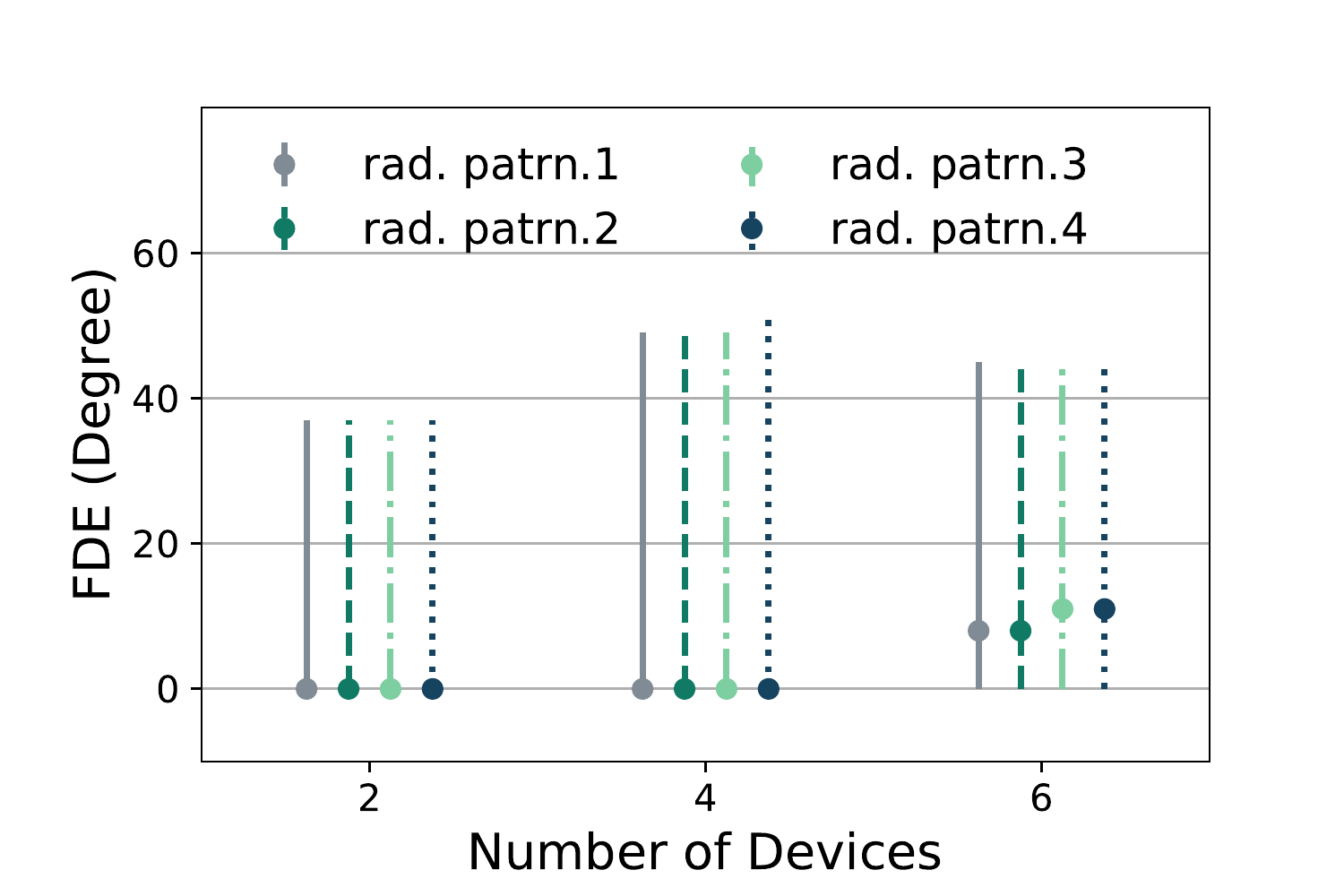}
            \caption{(a) Facing direction error grouped by $\widetilde{SNR}$. (b) 4 different radiation patterns, $2$ real and $2$ synthesized. (c) FDE when using each radiation pattern across all test cases.
            }            
            \label{fig:minis2}
        \end{minipage}%
        \vspace{-0.1in}
    \end{figure*}

    We find that when there are $4$ anchor devices, the median P2P estimation
    error is less than $50$ cm, which in turn results in marginal FDE error. 
    In other words, relaxing the assumption of known device location is indeed at a moderate cost.  
    \new

    $\blacksquare$ \textbf{User Mobility}: 
    We test the proposed algorithm under mobile scenarios, in which case the volunteer starts from an arbitrary location and takes a few random steps while speaking the voice command. 
    Table ~\ref{tab:mobileNmic} compares the median facing direction error (FDE) for mobile and static cases. 
	The result confirms that the maximum difference in median errors between static and mobile users is $11$ degrees.
	This is not surprising because human movements are slow, and in the time needed to just say ``Alexa'', the source shift is not excessive.
	This offers evidence that {\name} can reasonably handle user mobility.


\begin{table}[h]
        \begin{tabular}{c |c| c |c |c}
            \toprule
	{}&{Num. of}  & \textbf{Mobile User}  &\textbf{Static User} &\textbf{Static User} \\
	{}&{Devices } &6 Mics  &6 Mics & 4 Mics   \\\midrule \midrule 
	\multirow{3}{*}{\rotatebox[origin=c]{90}{FDE}}
	&2   &  $0^{\circ}$   & $0^{\circ}$  & $0^{\circ}$ \\
	&4   &  $11^{\circ}$  & $0^{\circ}$  & $20^{\circ}$  \\
	&6   &  $15^{\circ}$  & $8^{\circ}$ & $37^{\circ}$ \\\midrule
	\multirow{3}{*}{\rotatebox[origin=c]{90}{Loc. Err.}}
	&2   &66cm &68cm   &190cm \\
	&4   &59cm &49cm   &153cm \\
	&6   &51cm &37cm   &143cm    \\\bottomrule
           
        \end{tabular}
        \vspace{0.5em}
      \caption{Median FDE and localization error for static and mobile user, and varying microphones.}
        \label{tab:mobileNmic}
	\vspace*{-1.2em}
\end{table}

$\blacksquare$ \textbf{Varying Number of Microphones}:
The right $2$ columns of Table \ref{tab:mobileNmic} shows the median FDE and median localization error across $4$ and $6$ microphones.
This shows a significant change in error, e.g., with $6$ devices, the median FDE jumps to $40^{\circ}$.
This reveals the importance of diversity combining -- fewer microphones per device affects both the local and global diversity combining modules.
The AoA accuracy also degrades and this error further percolates through the system.
If marginal costs of adding more microphones is small, we believe this is certainly the right direction for maximal performance improvement.
\new

%

$\blacksquare$ \textbf{Varying $\widetilde{SNR}$}:
Figure ~\ref{fig:minis2}(a) plots the mean FDE under different Signal-to-Noise Ratio ($\widetilde{SNR}$).
The $\widetilde{SNR}$ indicates $\frac{signal+noise}{noise}$, as opposed to $\frac{signal}{noise}$.
This is because when the source signal is unknown, it is difficult to estimate SNR.
In our case, $\widetilde{SNR}$ varies mainly due to (a) the ambient noise in the room, and (b) the loudness of the voice.
We find that the performance of {\name} degrades gracefully with $\widetilde{SNR}$ except when the $\widetilde{SNR}$ becomes extremely low. 
This is evidence of reasonably robust channel estimation, as well as the benefits of radiation pattern matching.
\new


$\blacksquare$ \textbf{Radiation Pattern Template}:
During deployment, the radiation pattern of each user is simply not available.
Thus, {\name} uses a global radiation pattern, either from a specific user, or synthesized by averaging.
In our case, we carefully measure the radiation pattern of $2$ users (via multipath free experiments in anechoic-like chambers with nearby microphones around the user).
We also generate the average of these $2$ radiation patterns, and a final $4^{th}$ one that is deliberately distorted.
Figure \ref{fig:minis2}(b) plots all these $4$ voice radiation patterns.
To test robustness, we run FDE results by using each of the radiation patterns across all the test cases.
\new

Figure \ref{fig:minis2}(c) reports the FDE results, namely the median along with error bars at $25$ and $75$ percentile.
Evidently, results are robust, even when the distorted radiation pattern is used.
This suggests that the rough trend of the patterns matter during correlation; the precise values are not as critical.
This offers confidence that customization for radiation patterns is not crucial for real-world operation.
\new

%

    $\blacksquare$ \textbf{Computation Time}:
    Table ~\ref{tbl:comp} reports the median computation time for facing direction estimation and P2P localization.
    The median is $<0.5$s when executed on a Raspberry Pi 4 with quad core Cortex-A72 (ARM v8) 64-bit SoC @ 1.5GHz with 4GB  SDRAM. This is across all our reported results.
    Clearly, this is certainly within time budget, since completing a voice command is in the order of seconds.

\begin{table}[t]
        \begin{tabular}{c c c c}
            \toprule
            Num. of active devices & 2   & 4   & 6    \\ \midrule            \midrule
            P2P devices localization & {-}  & 578  & 631   \\ \hline
	    Facing device estimation &153 &305 & 451 \\ \bottomrule
        \end{tabular}
        \caption{Median computation time of \name\ in milliseconds using different number of active devices.}
        \vspace*{-2em}
        \label{tbl:comp}
    \end{table}
    

\section{Limitations and Discussions}



$\blacksquare$ \textbf{Single-microphone Devices:} 
A device $D_{s}$ may be equipped with a single microphone, hence cannot contribute its AoA for localization/triangulation.
However, so long as it has an embedded speaker, it can participate in P2P device localization, and its locations can be inferred.
Since user triangulation does not need all the devices, the user's location can also be computed without $D_{s}$'s location.
Now, when estimating the facing direction, it may be possible to identify that the user is facing $D_{s}$ from the correlation patterns.
We have left such analysis and algorithms to future work.
\new

$\blacksquare$ \textbf{Self Interference:} 
Many devices -- such as refrigerators, A/C machines, TVs, water faucets, etc. -- produce sounds as an outcome of their normal operation.
This sound will pollute the recorded voice signals and could cause the LoS power estimates to become erroneous.
We have not tested such cases in this paper.
However, given these sounds are \hl{typically stationary random processes}, their power profile may not change significantly.
Thus, it should be possible to characterize this noise floor (before the start of the command) and subtract it from the estimated LoS signal.
\new


$\blacksquare$ \textbf{Online learning}:
Many locations and surroundings of the devices may remain unchanged over time, suggesting strong channel coherence between two voice commands (from the same location). 
The {\name} system should learn from the past results and refine both user location and channel estimates. 
This work leaves this opportunity untapped.
\new


$\blacksquare$ \textbf{Leverage the full voice command}:
We have used only the short wake-up word to localize and infer the facing direction. 
While the user is speaking the rest of the voice command, we utilize that time to complete the facing direction computation.
If devices are more powerful, a longer part of the voice command could be used.
Longer signals (over stable channels) will help average out the white-sense stationery noises, likely refining the channel estimation results.


\section{Related Work}

\textbf{Inferring Pose and Facing Direction.} 
We believe this paper may be among the early works towards inferring facing direction from (unknown) human voice signals. 
Clearly, much work exists on voice radiation and in using acoustics for ranging \cite{peng2007beepbeep}, localization, motion tracking \cite{mao2019rnn,mao2016cat,smith2004tracking, yun2017strata, zhang2012swordfight}, etc. 
Radiation patterns (of antennas) have also been used for RF localization and sensing\cite{xiong2013arraytrack, wu2012csi, bocca2013multiple, kotaru2015spotfi, wang2017mfdl, karanam2018wifiTrack, sen2013avoiding, mariakakis2014sail, zheng2014CSI, vasisht2016decimeter}, however, source signals are often known in such systems.
Facing direction is difficult due to unknown source signals that vary slightly over adjacent facing directions.
\new

\textbf{AoA Estimation and Beamforming.}
Literature in beamforming and AoA is extremely mature~\cite{dokmanic2013roomshape, dokmanic2015thesis, ribeiro2010turning,anManocha2018reflection, anManocha2019diffraction}. 
For AoA detection one of the most commonly used algorithm is GCC-PHAT~\cite{knapp1976generalized, brandstein1997robust, benesty2004time, freire2011doa, van2012time}.
Recent years have also seen subspace based algorithms (e.g., MUSIC and ESPRIT) for AoA estimation \cite{roy1989esprit, schmidt1986music, sit2012direction, tang2014doathesis, xu2017weighted},
although they assume multiple uncorrelated source signals. 
Acoustic AoA and filtering have also been used in new types of imaging applications \cite{mao2018aim,mao2019mobile}. 
{\name} borrows from this literature to bring together the complete system -- we do not contribute here.
\new

\textbf{Acoustic Localization and Sensing.}
User localization using acoustic signals have also been well explored in the literature. 
Various approaches have been studied, including elapsed time of arrival~\cite{peng2007beepbeep}, 
exploiting wide-band signal and microphone array \cite{dokmanic2013roomshape, dokmanic2015thesis}, 
location estimation from voice commands~\cite{shen2020voloc}, and 
room shape determination from controlled acoustic transmissions~\cite{dokmanic2013roomshape, dokmanic2015thesis, ribeiro2010turning, anManocha2018reflection, anManocha2019diffraction}. 
Facing direction is a somewhat complementary problem; localization is necessary but not sufficient to solve facing direction.
\new

\textbf{Blind Channel Inference (BCI) and De-reverberation.} 
BCI a well known approach in wireless communication, with 
statistical methods such as second-order \cite{tong1994blind} and cyclostationary \cite{tong1991new} statistics, or sub-space based methods \cite{muquet2002subspace}.
Also, work on speech de-reverberation has proposed methods that estimates the source signal using linear prediction model, which employs gradient descend on the channel \cite{gillespie2001speech}.
Such techniques, while applicable in principle, are impractical due to the time and CPU budget in voice assistants.
{\name}'s contribution is in operating at the intersection of unique application specific constraints and opportunities.

\section{Conclusion}
Motivated by increasing number of voice assistants in homes and offices, we explore the problem of detecting the target device for a human voice command.  
The core technical challenge boils down to isolating the LoS power in a received (multipath-polluted) voice signal.
While the general LoS separation problem is extremely difficult, we find that the application presents opportunities.
This paper develops a solution using an iterative diversity combining algorithm and builds a fully functional system prototype.
While improvements are possible, we believe this is important step towards a real problem in the near future.

\bibliographystyle{ACM-Reference-Format}
\balance
\bibliography{reference}
\newpage

\end{document}